\documentclass[pre,aps,amsmath,amssymb,amsfonts,floatfix,superscriptaddress,twocolumn,footinbib]{revtex4-1}

\usepackage{graphicx}
\usepackage{amsmath,amsfonts}
\usepackage{xcolor}
\usepackage{color}
\usepackage{mathtools}
\usepackage{eufrak}
\usepackage{pgfplots}
\usepackage{multirow}

\newcommand{\up}{\uparrow}
\newcommand{\dn}{\downarrow}
\newcommand{\kv}{\ensuremath{\mathbf{k}}}
\newcommand{\qv}{\ensuremath{\mathbf{q}}}
\newcommand{\zerov}{\ensuremath{\mathbf{0}}}

\newcommand{\ch}{\ensuremath{\text{ch}}}
\newcommand{\sz}{\ensuremath{\text{sp}}}
\newcommand{\estimates}{\overset{\scriptscriptstyle\wedge}{=}}

\usepackage{tikz}
\usetikzlibrary{calc}
\usetikzlibrary{decorations.pathmorphing}
\usetikzlibrary{decorations.pathreplacing}
\usetikzlibrary{decorations.markings}
\usetikzlibrary{shapes.misc}
\usetikzlibrary{shapes}
\usetikzlibrary{positioning}
\usetikzlibrary{snakes}
\usetikzlibrary{arrows}
\usetikzlibrary{fadings}
\tikzfading[name=fade out,
inner color=transparent!0,
outer color=transparent!100]
\usetikzlibrary{calc,arrows}

\tikzset{snake it/.style={decorate, decoration=snake}}
\usetikzlibrary{3d}
\usepackage{mathtools}
    \tikzset{
            partial ellipse/.style args={#1:#2:#3}{
                        insert path={+ (#1:#3) arc (#1:#2:#3)}
                            }
                        }

\usetikzlibrary{calc}
\tikzset{
            inertial frame/.style = {x={(-20:2cm)}, y={(-160:2cm)}, z={(90:2cm)}},
              local frame/.style = {shift={(local origin)}, x={(40:.7cm)}, y={(150:.7cm)}, z={(105:.7cm)}}
          }

    \tikzset{middlearrow/.style={
                decoration={markings,
                            mark= at position 0.5 with {\arrow{#1}} ,
                                    },
                                            postaction={decorate}
                                                }
                                                }
\tikzset{cross/.style={cross out, draw, 
         minimum size=2*(#1-\pgflinewidth), 
                  inner sep=0pt, outer sep=0pt}}

\def\presuper#1#2%
  {\mathop{}%
   \mathopen{\vphantom{#2}}^{#1}%
   \kern-0.5\scriptspace%
   #2}

\usepackage{hyperref}
\hypersetup{colorlinks=true,breaklinks,linkcolor=blue,urlcolor=blue,citecolor=blue}
\begin{document}

    \pgfmathdeclarefunction{gauss}{2}{%
          \pgfmathparse{1/(#2*sqrt(2*pi))*exp(-((x-#1)^2)/(2*#2^2))}%
          }
    \pgfmathdeclarefunction{mgauss}{2}{%
          \pgfmathparse{-1/(#2*sqrt(2*pi))*exp(-((x-#1)^2)/(2*#2^2))}%
          }
    \pgfmathdeclarefunction{lorentzian}{2}{%
        \pgfmathparse{1/(#2*pi)*((#2)^2)/((x-#1)^2+(#2)^2)}%
          }
    \pgfmathdeclarefunction{mlorentzian}{2}{%
        \pgfmathparse{-1/(#2*pi)*((#2)^2)/((x-#1)^2+(#2)^2)}%
          }

\author{Friedrich Krien}
\email{fkrien@sissa.it}
\affiliation{International School for Advanced Studies (SISSA), Via Bonomea 265, 34136 Trieste, Italy}

\author{Erik G. C. P. van Loon}
\affiliation{Radboud University, Institute for Molecules and Materials, NL-6525 AJ Nijmegen, The Netherlands}

\author{Mikhail I. Katsnelson}
\affiliation{Radboud University, Institute for Molecules and Materials, NL-6525 AJ Nijmegen, The Netherlands}

\author{Alexander I. Lichtenstein}
\affiliation{Institute of Theoretical Physics, University of Hamburg, 20355 Hamburg, Germany}

\author{Massimo Capone}
\affiliation{International School for Advanced Studies (SISSA), Via Bonomea 265, 34136 Trieste, Italy}
\affiliation{CNR-IOM Democritos, Via Bonomea 265, 34136 Trieste, Italy}

%\pacs{
%71.30.+h%Metal-insulator transitions and other electronic transitions
%71.10.-w,%Theories and models of many-electron systems
%71.10.Fd,%Lattice fermion models (Hubbard model, etc.)
%71.27.+a,%Strongly correlated electron systems; heavy fermions
%71.45.Gm,%Exchange, correlation, dielectric and magnetic response functions, plasmons
%71.28.+d,%Narrow-band systems; intermediate-valence solids
%}

\title{Two-particle Fermi liquid parameters at the Mott transition: Vertex divergences, Landau parameters, and incoherent response in dynamical mean-field theory}

\begin{abstract}
    We consider the interaction-driven Mott transition at zero temperature from the viewpoint of microscopic Fermi liquid theory.
    To this end, we derive an exact expression for the Landau parameters within the dynamical mean-field theory (DMFT) approximation to the single-band Hubbard model.
    At the Mott transition the symmetric and the anti-symmetric Landau parameter diverge.
    The vanishing compressibility at the Mott transition directly implies the divergence of the forward scattering amplitude in the charge sector,
    which connects the proximity of the Mott phase to a tendency towards phase separation.
    We verify the expected behavior of the Landau parameters in a DMFT application to the Hubbard model on the triangular lattice at finite temperature.
    Exact conservation laws and the Ward identity are crucial to capture vertex divergences related to the Mott transition.
    We furthermore generalize Leggett's formula for the static susceptibility of the Fermi liquid to the static fermion-boson response function.
    In the charge sector the limits of small transferred momentum and frequency of this response function commute at the Mott transition.
\end{abstract}

\maketitle

\section{Introduction}

The Landau theory of Fermi liquids provides a fundamental phenomenological description of metals in their normal state~\cite{Landau56}. 
The theory accounts for (strong) interactions between the original fermions by introducing the concept of  quasi-particles~\cite{Woelfle18}, effective low-energy fermionic excitations which are characterized by an effective mass resulting from the interactions and by residual effective interactions.
Fermi liquid theory is applicable as long as the interacting system is continuously connected to the non-interacting fermion gas,
that is, no phase transition occurs. The theory makes general statements about the physical properties of Fermi liquids, which can be directly connected with experiments.
However, the values of the quasi-particle effective mass and the Landau parameters describing the residual interactions between quasi-particles must be either derived from a microscopic theory of a well-defined model, or extracted from experiments. In this work we focus on the former strategy, addressing the Landau theory for the Mott-Hubbard transition as described by the single-band Hubbard model.

A semi-phenomenological way to obtain non-perturbative numerical results for the Landau parameters in variational Monte Carlo studies is to fit the energy of low-lying particle-hole excitations with the Fermi liquid energy functional~\cite{Kwon94,Lee18}.
On the other hand, analytical expressions for the Landau parameters from first principles are frequently obtained by means of diagrammatic perturbation theory around the non-interacting limit, see, for example,~\cite{Fuseya00,Frigeri02,Chubukov18}. However, perturbation theory can not capture the breakdown of the Fermi liquid picture at an interaction-driven metal-to-insulator transition. 

A way to derive a microscopic Landau theory is to solve the Hubbard model using the variational Gutzwiller approximation~\cite{Gutzwiller63}, or the equivalent Kotliar-Ruckenstein slave-boson mean-field~\cite{Kotliar86,Li94}.
These methods describe a strongly renormalized, almost localized Fermi liquid and its disappearance at the Mott transition~\cite{Vollhardt84}.
The behavior of the Landau parameters close to the metal-insulator transition is especially interesting:
At the critical interaction the symmetric Landau parameter $\mathfrak{f}^{\,\ch}$ diverges~\cite{Vollhardt84,Fresard12}, in correspondence to the charge localization,
whereas the anti-symmetric one $\mathfrak{f}^{\,\sz}$ remains finite. Here the labels '$\ch$' and '$\sz$' indicate correspondence to the charge and spin sector, respectively.

On the other hand, when a Landau parameter $\mathfrak{f}$ approaches the value $-1$, in general~\cite{Kiselev17} a Pomeranchuk instability occurs,
which can be favored by non-local interactions~\cite{Lhoutellier15}.
The symmetric Landau parameter of a multi-orbital Hubbard model in the so-called Hund's metal regime has
recently been calculated using the slave-spin method~\cite{deMedici05,deMedici17},
which predicts a phase separation as an instability of the Mott insulator upon doping~\cite{deMedici17-2} which takes place
just above the critical interaction strength for the Mott transition.
The instability indeed tracks the evolution of the critical interaction as a function of different control parameters~\cite{Arribi18},
establishing a direct link between the Mott transition and the phase separation. 
Previous studies suggest a similar scenario also for the single-band Hubbard model~\cite{Furukawa91,Furukawa93,Nourafkan19}.

The development of the dynamical mean-field theory~\cite{Metzner89,Georges96} (DMFT) has widened our understanding of the Mott metal-insulator transition in the Hubbard model extending the previous results within a non-perturbative and conserving approach.
DMFT describes the evolution from the metal to the Mott insulator in terms of the reduction and the vanishing of the quasi-particle weight $Z$,
which within DMFT coincides with the inverse of the effective mass enhancement $Z = m/m^*$, one of the main parameters of the Fermi liquid theory.
However, while a Fermi liquid picture of this Mott metal-insulator transition was developed~\cite{Kotliar99,Kotliar00} in terms of $Z$, surprisingly
little is known about the Landau parameters in DMFT, despite their central role for the theory of Fermi liquids. A notable exception is Ref. \cite{Capone02}, where a Landau approach has been used to estimate the Cooper instability in a multi-band Hubbard model. The present work fills this gap with a thorough investigation of the Landau parameters in the single-band Hubbard model with a special focus on the approach to the interaction-driven Mott-Hubbard transition.

The Landau parameters have a crucial physical significance, as they may be interpreted as the residual interaction between the quasi-particles~\cite{Landau80,Noziere97,Mora15}.
From a technical point of view, the interaction character implies that they are two-particle quantities.
In particular, as we will detail in the following, in a microscopic Fermi liquid theory the Landau parameters are given by the dynamic limit of the two-particle vertex function, $\mathfrak{f}\propto \presuper{0\!}FZ^2$.
The dynamic limit $\presuper{0\!}F$ of the vertex corresponds to the forward-scattering limit of vanishing momentum and frequency transfer,
where first $\qv\rightarrow\mathbf{0}$ and then $\omega\rightarrow0$, respectively,
and it captures all forward scatterings \textit{except} those between quasi-particles and quasi-holes.
On the other hand, the static limit $\presuper{\infty\!}F$, where first $\omega\rightarrow0$ then $\qv\rightarrow\mathbf{0}$,
accounts for \textit{all} forward scatterings of particles and holes, and it is the quantity responsible of actual physical instabilities.
For this reason $\presuper{\infty\!}F$, rather than $\presuper{0\!}F$, is often denoted as the forward scattering amplitude~\cite{Noziere97}.

In general it is difficult to calculate the vertex function, due to its dependence on three real frequencies and momenta.
In an isotropic system, such as $^3$He, the Landau parameters can be expanded into the Legendre
polynomials. This simplification leads to several prominent Fermi liquid relations for the isotropic Fermi liquid,
such as Leggett's formula for the static susceptibility~\cite{Legget65,Kiselev17,Wu18,Chubukov18}.
On the other hand, in a spatially inhomogeneous system like a lattice model the Landau
parameters acquire rich momentum dependence which emerges already at the second order of perturbation
theory~\cite{Fuseya00,Frigeri02}. This may appear as a serious obstacle to compute the Landau parameters in the non-perturbative
DMFT, which is defined for lattice systems. As will be shown in this work, this is not the case, the Landau parameters can be calculated easily in DMFT.

In fact, recent technical advancements have eased the direct calculation of the vertex function:
DMFT maps the lattice model onto an Anderson impurity model whose hybridization function must be determined
self-consistently as we briefly recall below. Therefore, the DMFT evaluation of the vertex function requires to
compute four-point correlation functions of the impurity model. This can be done using continuous-time quantum
Monte Carlo (CTQMC) solvers~\cite{Rubtsov05,Werner06,Gull11} with improved estimators~\cite{Hafermann12,Hafermann14,Gunacker16},
which allow the measurement of impurity vertices at finite temperature with very high accuracy.
In the calculation of the DMFT susceptibility the numerical error can be reduced even further,
by separating it into local and non-local contributions~\cite{Pruschke96,Rubtsov12,vanLoon15,vanLoon16,Krien19} and
by taking vertex asymptotics into account~\cite{Kunes11,Wentzell16}. These improvements have given rise to diagrammatic extensions
of DMFT~\cite{[{For a recent review, see: }]Rohringer17} and opened a window into the two-particle level of its impurity model, which led to the discovery of
vertex divergences~\cite{Schaefer13}. Some divergences of the impurity vertex function have been related to the
Mott transition~\cite{Rohringer12,vanLoon18}. Furthermore, the two-particle self-energy $\gamma$
(irreducible vertex) also shows divergences, sometimes located in the vicinity of the Mott insulator~\cite{Schaefer13},
but they also arise in systems without Mott localization~\cite{Chalupa18} and even in the atomic limit~\cite{Thunstroem18}.
Divergences of the two-particle self-energy have been related to the multi-valuedness of the Luttinger-Ward functional~\cite{Kozik15,Gunnarsson17,Vucicevic18}.

A natural question is whether some vertex divergences can explain characteristic features of the Mott transition.
For example, it was hypothesized in Ref.~\cite{Chitra01} that in the Mott phase there may exist a divergent scattering amplitude in the charge sector.
If this prediction was confirmed in a microscopic scheme which properly accounts for the Mott transition,
it would strengthen the case for the somewhat counter-intuitive tendency towards phase separation --associated with a divergent compressibility-- close a Mott insulator,
where the same compressibility must vanish. In the two-dimensional Hubbard model phase separation close to the Mott transition
has been widely debated~\cite{Furukawa91,Furukawa93,Cosentini98,Sorella15} also as a possible source of charge-ordering instabilities~\cite{Caprara17}
or even superconductivity~\cite{Grilli95,Emery93}.
A finite-temperature divergent compressibility in DMFT has been suggested to underlie the $\alpha-\gamma$ transition in cerium~\cite{Kotliar02}. 

Another question which arises is the connection of divergences of the impurity vertex of DMFT to physical instabilities of the lattice model:
Despite a divergence of the two-particle self-energy $\gamma$ the impurity vertex function $f$ may remain finite.
And, in turn, a divergence of $f$ does not always imply a divergence in the DMFT approximation to the lattice vertex function $F$.
For example, it is widely believed that despite the divergence of the impurity spin vertex in the Mott phase at zero temperature
the uniform spin susceptibility of DMFT remains finite, due to the effective exchange $J=\tilde{t}^{\,2}/U$~\cite{Rozenberg94,Georges96,Hewson16}. 
In fact, as will be shown in this work, even the divergence of the lattice forward scattering amplitude $\presuper{\infty\!}F$ does not always imply the divergence of a susceptibility.

The aim and outline of this work are as follows:
In Sec.~\ref{sec:flt0} we recollect the main ingredients of the Fermi liquid theory,
including the Landau parameters, the Ward identities for the fermion-boson response,
and Leggett's decomposition of the static susceptibility~\cite{Legget65}.
Next, we apply the DMFT approximation in Sec.~\ref{sec:dmft} and obtain a complete set of Fermi liquid relations,
whose behavior we analyze in the limit of vanishing quasi-particle weight.
This shows that the static and dynamic limit of the charge fermion-boson response commute at the Mott transition and several vertex functions diverge.
One divergent vertex is the charge forward scattering amplitude,
which implies that a Fermi liquid with quasi-particles of nearly divergent effective mass exhibits a tendency towards the phase separation.
In Sec.~\ref{sec:mott} we report DMFT results for the Hubbard model on the triangular lattice at finite temperature,
which confirm several of the vertex divergences.
We present numerical results for the fermion-boson response and explain its qualitative behavior near the critical interaction of the Mott transition.
We summarize and discuss our results in Sec.~\ref{sec:discussion} and close with the conclusions in Sec.~\ref{sec:conclusions}.

\section{Fermi liquid theory at $T=0$}\label{sec:flt0}
In this section we provide a self-contained and pedagogical recollection of the microscopic Landau Fermi liquid theory.
Readers may jump to Secs.~\ref{sec:flparms} and~\ref{sec:z0} as well as Table~\ref{tab:scaling} to find our main results.
While the derivations are general we focus for concreteness on the single-band Hubbard model on the triangular lattice,
\begin{align}
    H = &-\sum_{\langle ij\rangle\sigma}\tilde{t}_{ij} c^\dagger_{i\sigma}c^{}_{j\sigma}+ U\sum_{i} n_{i\up} n_{i\dn},\label{eq:hubbard}
\end{align}
where $\tilde{t}_{ij}$ is the isotropic nearest neighbor hopping between lattice sites $i,j$ .
We use the hopping amplitude $\tilde{t}=1$ as the unit of energy. $c^{},c^\dagger$ are the annihilation and construction operators, $\sigma=\up,\dn$ the spin index.

The starting point for the microscopic Fermi liquid theory are the analytical properties of the \textit{causal} Green's function
$G^c_{\kv\sigma}(t-t')=-\imath\langle T_t c_{\kv\sigma}(t)c^\dagger_{\kv\sigma}(t')\rangle$,
which is used in perturbation theory for real times $t,t'$~\cite{Noziere97}, where $T_t$ is the time-ordering operator.
The spin label $\sigma$ will be dropped where unambiguous.
The frequency transform of this function can be expressed in the following way [cf. Appendix~\ref{app:gf}],
\begin{align}
    G^c_{\kv\nu}=& n_f(-\nu)G^r_{\kv\nu}+n_f(\nu) G^a_{\kv\nu}.\label{eq:gcgr}
\end{align}
Here, $\nu$ is the real frequency, $n_f(\nu)=(1+e^{\beta\nu})^{-1}$ is the Fermi function, $G^r$ and $G^a$ are the retarded and advanced Green's functions.
The latter are analytical in the upper/lower complex half-plane, respectively, whereas $G^c$ itself is not analytical in either half-plane.

The causal Green's function~\eqref{eq:gcgr} has a hole-like (advanced) and a particle-like (retarded) component.
For finite temperature these components are mixed in the vicinity of the Fermi level $\nu=0$, due to the thermal softening of the Fermi function.
Here we focus on the zero temperature limit $T={\beta}^{-1}\rightarrow0$,
hence the Fermi function in Eq.~\eqref{eq:gcgr} becomes a Heaviside step-function, $n_f(\nu)\rightarrow\theta(-\nu)$.

\subsection{Fermi-liquid Green's function}
The central assumption of the Fermi liquid theory is that even in presence of an interaction the Green's function has a simple structure,
with a pole of weight $Z_\kv$ at the Fermi level.
In the neighborhood of the Fermi momentum, $\kv\approx\kv_F$, one may write the Green's function as
\begin{align}
    G^c_{\kv\nu}=&\frac{Z_\kv}{\nu-\tilde{\varepsilon}_{\kv}+\mu+\imath\eta}+G^{c,\text{inc}}_{\kv\nu}.\label{eq:gffl}
\end{align}
Here, $\tilde{\varepsilon}_\kv$ is the renormalized (quasi-particle) dispersion, $\mu$ is the chemical potential.
$G^{c,\text{inc}}$ is an incoherent background, by assumption a smooth function of $\kv$ and $\nu$.
$\eta=0^\pm$ is an infinitesimal number. The first term can be obtained from the generic expression of the Green's function as a function of the self-energy $\Sigma_{\kv}(\nu)$ expanding the latter around the Fermi level.
This defines the quasi-particle weight,
\begin{align}
  Z_\kv^{-1} = 1 - \left.\frac{\partial\Re\Sigma_\kv(\nu)}{\partial\nu}\right|_{\nu=0},\label{eq:qpweight}
\end{align}
and the quasi-particle dispersion,
\begin{align}
 \tilde{\varepsilon}_\kv-\mu=Z_\kv[\varepsilon_{\kv}-\mu+\Re\Sigma_{\kv}(0)].\label{eq:qpdispersion}
\end{align}
The zeros of $\tilde{\varepsilon}_\kv-\mu$ define the manifold of Fermi vectors $\kv_F$.
In combination with equation~\eqref{eq:gcgr} and $n_f(\nu)=\theta(-\nu)$ one sees that 
$G^c_{\kv\nu}$ has a pole of weight $Z_\kv$ in the lower complex half-plane ($\eta=0^+$, retarded) when $\kv$ lies outside of the Fermi surface.
This pole then represents a quasi-particle.
On the other hand, when $\kv$ lies within or on the Fermi surface, the pole is in the upper half-plane ($\eta=0^-$, advanced), representing a quasi-hole.

The label $c$ denoting the causal Green's function $G^c$ will be dropped in the remainder of this section.

\subsection{Discontinuity of the bubble}\label{sec:flt0:disc}
The formal derivation of the Fermi liquid theory following Landau, Nozi\`eres and Luttinger~\cite{Landau80,Noziere97,Abrikosov75,Noziere62-1,Noziere62-2} is obtained from the two-particle level,
by analysis of the analytical structure of the particle-hole spectrum.

The crucial point is that the pole structure of the causal Fermi liquid Green's function~\eqref{eq:gffl} leads to the counter-intuitive situation that the limits
$\qv\rightarrow\zerov$ and $\omega\rightarrow0$ of the bubble $G^2_{\kv\nu}(\qv,\omega)$ do not commute, where
\begin{align}
    G^2_{\kv\nu}(\qv,\omega)=G_{\kv-\frac{\qv}{2},\nu-\frac{\omega}{2}}\;G_{\kv+\frac{\qv}{2},\nu+\frac{\omega}{2}}.\label{eq:bubbledef}
\end{align}

Taking the homogeneous limit $\qv\rightarrow\zerov$ first, $\eta$ in Eq.~\eqref{eq:gffl} has the same sign for both Green's functions.
Therefore, $G_{\kv,\nu-\omega/2}$ and $G_{\kv,\nu+\omega/2}$ have their poles in the same complex half-plane.
These poles either represent two quasi-holes or two quasi-particles with the same momentum $\kv$, but never a quasi-particle-hole pair.
Taking the limit $\omega\rightarrow0$ {subsequently} does not change this situation.

However, when taking the limit $\omega\rightarrow0$ first, a peculiarity arises at the Fermi momentum $\kv=\kv_F$:
In the case that $\kv_F-\frac{\qv}{2}$ lies inside the Fermi surface, $\kv_F+\frac{\qv}{2}$ in general [cf. Appendix~\ref{app:limits}] lies outside and vice versa.
As a consequence, the poles of $G_{\kv_F-\qv/2,\nu}$ and $G_{\kv_F+\qv/2,\nu}$ describe a quasi-particle-hole pair.

These distinct limits of the bubble are defined as,
\begin{align}
    \presuper{\infty\!}G^2_{\kv\nu}\equiv&\lim\limits_{\qv\rightarrow\mathbf{0}}\lim\limits_{\omega\rightarrow0}G_{\kv-\frac{\qv}{2},\nu-\frac{\omega}{2}}G_{\kv+\frac{\qv}{2},\nu+\frac{\omega}{2}},\\
    \presuper{0\!}G^2_{\kv\nu}\equiv&\lim\limits_{\omega\rightarrow0}\lim\limits_{\qv\rightarrow\mathbf{0}}G_{\kv-\frac{\qv}{2},\nu-\frac{\omega}{2}}G_{\kv+\frac{\qv}{2},\nu+\frac{\omega}{2}},\label{eq:bubblelimits}
\end{align}
where the left superscript of $\presuper{\mathfrak{r}}G^2$ denotes the ratio $\mathfrak{r}=|\qv|/\omega$.
We will refer to $\mathfrak{r}=\infty$ and $\mathfrak{r}=0$ in the following as the static homogeneous and the dynamic homogeneous limit, respectively,
(we abbreviate as the static and the dynamic limit where unambiguous).

One further defines the discontinuity of the bubble as the difference between the static and the dynamic limit,
\begin{align}
    R_{\kv\nu}=&\presuper{\infty\!}G^2_{\kv\nu}-\presuper{0\!}G^2_{\kv\nu}=-2\pi\imath Z^2_{\kv}\delta(\nu)\delta(\tilde{\varepsilon}_\kv-\mu),\label{eq:rdef}
\end{align}
which has poles at $\kv=\kv_F$ and $\nu=0$ and is zero elsewhere.
The explicit expression for $R$ is derived in Ref.~\cite{Noziere97}, it is \textit{not} restricted to isotropic systems.

We comment on some technical difficulties concerning the proper definition of the static homogeneous limit in Appendix~\ref{app:limits}.

\subsection{Vertex function and Landau parameters}\label{sec:landauparm}
We introduce the vertex function $F^\alpha_{kk'q}$, where we use the short notation $k=(\kv,\nu)$ and $q=(\qv,\omega)$.
The label $\alpha$ denotes the charge ($\alpha=\ch$) and spin ($\alpha=x,y,z=\sz$) channels,
where the latter can be comprised into one index due to rotational invariance.
Fig.~\ref{fig:3leg} d) shows a diagrammatic representation of $F$ and the convention for its labels $k, k'$, and $q$ used in this text.

The vertex function $F$ is constructed from the bubble $G^2$ via the Bethe-Salpeter equation,
\begin{align}
    \presuper{\mathfrak{r}}F^\alpha_{kk'}=\Gamma^\alpha_{kk'}+\int_{k''}\Gamma^\alpha_{kk''}\presuper{\mathfrak{r}}G^2_{k''}\presuper{\mathfrak{r}}F^\alpha_{k''k'}.\label{eq:bselimits}
\end{align}
Here, $\Gamma^\alpha_{kk'q}$ is the two-particle self-energy, which is irreducible with respect to the bubble $G^2$.
The integral over $k''$ implies normalized summation/integration over $\kv''$ and $\nu''$.
For the Hubbard model~\eqref{eq:hubbard} we have $\int_{k}=\frac{1}{2\pi N}\sum_{\kv}\int_{-\infty}^{+\infty}d\nu$,
with $N$ the number of lattice sites.

In Eq.~\eqref{eq:bselimits} the double limit $q=(\qv,\omega)\rightarrow0$ has already been taken.
In fact, since $F$ is constructed from the bubble $G^2$, it inherits the ambiguity of this limit.
This means that $F$ and $G^2$ in Eq.~\eqref{eq:bselimits} both carry a label $\mathfrak{r}$, indicating that either the static or dynamic limit is taken.

On the other hand, in the Fermi liquid the limit $q\rightarrow0$ of the two-particle self-energy $\Gamma$ is \textit{not} ambiguous, (see, for example, Ref.~\cite{Noziere97}),
since by construction $\Gamma$ is free of the problematic bubble insertions $G^2$.
Hence, the homogeneous limit of the two-particle self-energy does not inherit a label $\mathfrak{r}$,
\begin{align}
\presuper{\mathfrak{r}}\Gamma_{kk'}\equiv\Gamma_{kk'}.
\end{align}
As a consequence, $\Gamma$ can be eliminated from Eq.~\eqref{eq:bselimits},
leading to an important exact relation between the static and dynamic limits of the vertex function,
\begin{align}
   \presuper{\infty\!}F^\alpha_{kk'}=\presuper{0\!}F^\alpha_{kk'}+\int_{k''}\presuper{0\!}F^\alpha_{kk''}R_{k''}\presuper{\infty\!}F^\alpha_{k''k'},
   \label{eq:bselimits_inv}
\end{align}
where $R$ is the discontinuity of the bubble defined in Eq.~\eqref{eq:rdef}.

We comment on the physical significance of the limits $\presuper{\infty\!}F$ and $\presuper{0\!}F$ of the vertex function and of Eq.~\eqref{eq:bselimits_inv}:

The static limit $\presuper{\mathfrak{r}=\infty}F$, the so-called forward scattering amplitude,
describes the physical situation of small momentum $\delta\qv\approx\zerov$ and strictly vanishing energy transfer $\omega=0$. 
This includes, but is not limited to, the scattering events between quasi-particles and quasi-holes that leave both of them on the Fermi surface.

On the other hand, the scatterings associated to $\presuper{\mathfrak{r}=0}F$ imply the situation of small energy
$\delta\omega\approx0$ and vanishing momentum transfer $\qv=\zerov$.
As explained in Sec.~\ref{sec:flt0:disc}, the peculiarity of this limit 
is precisely that it does \textit{not} account for quasi-particle-hole contributions.
Hence, $\presuper{0}F$ describes all forward scatterings except the ones between quasi-particles and quasi-holes
[such as incoherent-on-incoherent or coherent-on-incoherent scatterings].

The second term on the right-hand-side of Eq.~\eqref{eq:bselimits_inv} therefore represents
the contribution of quasi-particle-hole scatterings to the static limit $\presuper{\infty\!}F$.
The latter is recovered from the dynamic limit $\presuper{0\!}F$ by taking repeated scatterings of this type into account,
while $\presuper{0\!}F$ assumes the role of an effective quasi-particle interaction.
One defines $\mathfrak{f}^\alpha_{kk'}\propto Z_\kv Z_{\kv'}\presuper{0\!}F^\alpha_{kk'}$, the Landau parameters,
where $Z_\kv$ is the quasi-particle weight from Eq.~\eqref{eq:gffl} and $k, k'$ lie on the Fermi surface.
One often refers to $\mathfrak{f}^{\,\ch}$ and $\mathfrak{f}^{\,\sz}$ as the symmetric and anti-symmetric Landau parameter, respectively~\cite{Noziere97}.

\subsection{Three-leg vertex and Ward identity}\label{sec:wardidt0}
\begin{figure}
    \begin{center}
    \includegraphics[width=0.45\textwidth]{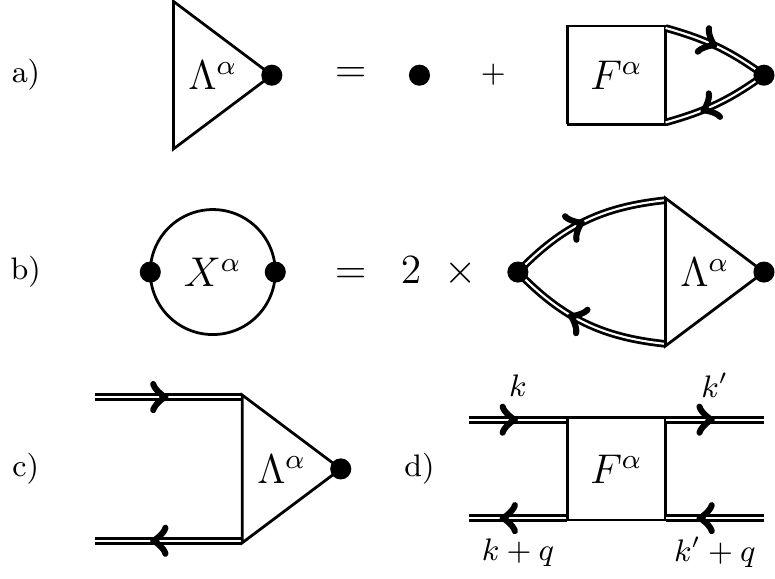}
    \end{center}
    \vspace{-.3cm}
    \caption{\label{fig:3leg} a) Diagrammatic representation of the three-leg vertex $\Lambda$ and its relation to the vertex function $F$.
    Filled circles imply summation over internal indices of incoming Green's function lines (double lines with arrows), which are all interacting.
    b) Relation between susceptibility (total response) $X$ and three-leg vertex $\Lambda$. c) Fermion-boson response function $L$. d) Label convention for $F$.}
    \end{figure}

We introduce a central object of this work, the three-leg vertex $\Lambda^\alpha_{kq}$.
The latter is obtained from the vertex function $F$ by attaching a bubble $G^2$ to $F$, closing the open ends,
and adding $1$, as in Fig.~\ref{fig:3leg} a),
\begin{align}
    \Lambda^\alpha_{kq} = 1 + \int_{k'}F^\alpha_{kk'q}G_{k'}G_{k'+q}\label{eq:lambdat0}.
\end{align}

Although $\Lambda$ itself may not have a direct physical interpretation, it is closely related to a physical response function,
the fermion-boson response function [see Fig.~\ref{fig:3leg} c) and Appendix~\ref{app:ac:g3}],
\begin{align}
    L^{\alpha}_{kq}=G_kG_{k+q}\Lambda^{\alpha}_{kq}.\label{eq:g3def}
\end{align}

In fact, $L_{kq}$ is best construed as the response of an electronic state with momentum and energy vector $k=(\kv,\nu)$,
which responds to an applied field with spatial and temporal dependence $q=(\qv,\omega)$.

In the limit $q\rightarrow0$ this can be seen using Ward's identity, which allows to calculate
$\presuper{\mathfrak{r}}L^{\alpha}_k=\presuper{\mathfrak{r}}G^2_k\;\presuper{\mathfrak{r}}\Lambda^\alpha_k$ explicitly,
where again $\mathfrak{r}$ indicates how the double limit is taken.
We show in Appendix~\ref{app:ac:w0} that one obtains the following relations for the static homogeneous limit,
\begin{align}
   \presuper{\infty\!}L^{\ch}_k=&-\frac{dG_{k}}{d\mu},\label{eq:ward:g3statch}\\
   \presuper{\infty\!}L^{\sz}_k=&-\frac{dG_{k\uparrow}}{dh},\label{eq:ward:g3statsp}
\end{align}
where on the right-hand-sides appear derivatives with respect to the chemical potential $\mu$ and the homogeneous magnetic field $h$ directed along the $z$-axis.

The Ward identities~\eqref{eq:ward:g3statch} and~\eqref{eq:ward:g3statsp} for $\presuper{\infty\!}L$ have a straightforward physical interpretation:
Upon a small change of the chemical potential $\delta\mu$ or magnetic field $\delta h$, within the linear response regime,
the spectral weight of electronic states with momentum $\kv$ and energy $\nu$ is changed by an
amount $-\delta\mu\presuper{\infty\!}L^{\ch}_k$ and $-\delta h\presuper{\infty\!}L^{\sz}_k$, respectively, (see also Sec.~\ref{sec:fbresponse} and Ref.~\cite{vanLoon18}).
The response function $L$ is therefore more rich in information than the susceptibility $X^\alpha_q=2\int_k L^{\alpha}_{kq}$,
which merely describes the total response of the electronic spectrum.
The relation between $X$ and $\Lambda$ is depicted diagrammatically in Fig.~\ref{fig:3leg} b).

For the dynamic limit of $L$, on the other hand, one finds the following relation [see Appendix~\ref{app:ac:ward}, Eq.~\eqref{app:wardidcausal}],
\begin{align}
   \presuper{0\!}L^{\alpha}_k=&-\frac{dG_{k}}{d\nu}.\label{eq:ward:g3dyn}
\end{align}
Note that this relation is valid for $\alpha=\ch,\sz$, leading to the same right-hand-side.
A physical interpretation of Eq.~\eqref{eq:ward:g3dyn} is less obvious than for the static limit $\presuper{\infty\!}L$.
The significance of the dynamic limit $\presuper{0\!}L$ will be articulated over the course of this work.

The Ward identities~\eqref{eq:ward:g3statch}-\eqref{eq:ward:g3dyn} can be reformulated in terms of the three-leg vertex $\Lambda$ via Eq.~\eqref{eq:g3def},
and by making use of Dyson's equation $G^{-1}_{k\sigma}=\nu-\varepsilon_\kv+\sigma h+\mu-\Sigma_{k\sigma}+\imath\eta$, where $\Sigma$ is the electronic self-energy.
The result is,
\begin{align}
    \presuper{\infty\!}\Lambda^\alpha_{k}=&\begin{cases}
        1-\frac{d\Sigma_{k}}{d\mu} & \text{for}\;\alpha=\ch, \\
        1-\frac{d\Sigma_{k\uparrow}}{dh} & \text{for}\;\alpha=\sz,
    \end{cases}\label{eq:lambdastatt0}\\
    \presuper{0\!}\Lambda^\alpha_{k}=&1-\frac{d\Sigma_k}{d\nu}.\label{eq:lambdadynt0}
\end{align}
We note that if $k=k_F$ lies at the Fermi level Eq.~\eqref{eq:lambdadynt0} relates the dynamic limit of the three-leg vertex to the quasi-particle weight
[cf. Eq.~\eqref{eq:qpweight}], $\presuper{0\!}\Lambda^\alpha_{\kv=\kv_F,\nu=0}=Z^{-1}_{\kv_F}$.

\subsection{Leggett's decomposition}\label{sec:legget}
We discuss the relation between the static and dynamic limits of $\Lambda$ and $L$, respectively.
We also do this for the susceptibility $X$, which recovers a result of Leggett~\cite{Legget65}.

First, we recall that the static and dynamic homogeneous limits of the vertex function $F$ are related via Eq.~\eqref{eq:bselimits_inv}.
From that relation follows a similar one for the three-leg vertex $\Lambda$ (see Refs.~\cite{Noziere97,Landau80} and Appendix~\ref{app:decomp}),
\begin{align}
    \presuper{\infty\!}\Lambda^\alpha_{k}=\presuper{0\!}\Lambda^\alpha_{k}+\int_{k'}\presuper{0\!}F^\alpha_{kk'}R_{k'}\presuper{\infty\!}\Lambda^\alpha_{k'},\label{eq:fbvertex}
\end{align}
which, in fact, gives rise to the linearized Boltzmann equation (or Landau's kinetic equation)~\cite{Noziere97}.

In Appendix~\ref{app:decomp} we show that from Eq.~\eqref{eq:fbvertex} follows also a relation between $\presuper{\infty\!}L$ and $\presuper{0\!}L$,
\begin{align}
    \presuper{\infty\!}L^{\alpha}_{k}=\presuper{0\!}L^{\alpha}_{k}+\int_{k'}\left(\delta_{kk'}+\presuper{0\!}G^2_{k}\;\presuper{0\!}F^\alpha_{kk'}\right)R_{k'}\presuper{\infty\!}\Lambda^\alpha_{k'},\label{eq:g3decomp}
\end{align}
where $\delta_{kk'}$ implies a factor $2\pi N$.
The integral of $\presuper{\infty\!}L$ yields the total static response, that is, the static homogeneous susceptibility,
$\presuper{\infty\!}X^\alpha=2\int_k \presuper{\infty\!}L^{\alpha}_k$,
\begin{align}
    \presuper{\infty\!}X^\ch=-\imath\frac{d\langle n\rangle}{d\mu},\;\;\;\presuper{\infty\!}X^\ch=-\imath\frac{d\langle m\rangle}{dh},\label{eq:suscs}
\end{align}
where $\langle n\rangle=\langle n_\up\rangle+\langle n_\dn\rangle$ and $\langle m\rangle=\langle n_\up\rangle-\langle n_\dn\rangle$ denote
the total density and magnetization per site, respectively.
The factor $\imath$ originates from the causal Green's function~\eqref{eq:gcgr} [cf. Appendix~\ref{app:gf}].

Performing the integration in Eq.~\eqref{eq:g3decomp} leads to,
\begin{align}
    \presuper{\infty\!}X^\alpha=\presuper{0\!}X^\alpha+2\int_{k'}\presuper{0\!}\Lambda^\alpha_{k'}\;R_{k'}\presuper{\infty\!}\Lambda^\alpha_{k'},\label{eq:legget}
\end{align}
where we have identified the three-leg vertex $\presuper{0\!}\Lambda^\alpha_k=\int_k(\delta_{kk'}+\presuper{0\!}G^2_{k}\;\presuper{0\!}F^\alpha_{kk'})$
using Def.~\eqref{eq:lambdat0} \footnote{This is a 'left-handed' three-leg vertex,
with the tapered Green's function legs in Fig.~\ref{fig:3leg} a) pointing to the left.
In the limit $q\rightarrow0$ it is equivalent to the 'right-handed' vertex due to the crossing symmetry.}
and the dynamic homogeneous susceptibility, $\presuper{0\!}X^\alpha=2\int_k\presuper{0\!}L^{\alpha}_k$.

However, $\presuper{0\!}L^{}$ does not contribute to the static susceptibility $\presuper{\infty\!}X$ in Eq.~\eqref{eq:legget}.
This can be seen using the Ward identity~\eqref{eq:ward:g3dyn}. 
The frequency integral over $\nu$ vanishes, $\presuper{0\!}X^\alpha=2\int_k\presuper{0\!}L^{\alpha}_k=-2\int_k\frac{dG}{d\nu}=0$,
since Green's function is zero at the boundaries $\pm\infty$. Physically this is a consequence of total charge and spin conservation.
Therefore, the entire static susceptibility $\presuper{\infty\!}X$ is given by the remainder on the right-hand-side of Eq.~\eqref{eq:legget}.

Lastly, we show that Eq.~\eqref{eq:legget} leads to a decomposition of the susceptibility due to Leggett.
Solving Eq.~\eqref{eq:fbvertex} for $\presuper{\infty\!}\Lambda$ via matrix inversion and inserting the result into Eq.~\eqref{eq:legget}
we can bring the latter into the following form,
\begin{align}
    \presuper{\infty\!}X^\alpha=2\iint_{k,k'}\presuper{0\!}\Lambda^\alpha_kR_k\left(\delta_{kk'}-\presuper{0\!}F^\alpha_{kk'}R_{k'}\right)^{-1}\presuper{0\!}\Lambda^\alpha_{k'},\label{eq:legget2}
\end{align}
where the inverse indicates a matrix inversion with respect to the indices $k$ and $k'$.
In Eq.~\eqref{eq:legget2} we have already omitted the vanishing $\presuper{0\!}X^\alpha$.

The static susceptibility $\presuper{\infty\!}X$ is therefore determined entirely by the quantities $R$, $\presuper{0\!}\Lambda$, and $\presuper{0\!}F$.
Inserting Eq.~\eqref{eq:rdef} for the discontinuity $R$ into equation~\eqref{eq:legget2} one is left with integrals over the Fermi surface and
the Ward identity~\eqref{eq:lambdadynt0} can be applied, $\presuper{0\!}\Lambda^\alpha_{k_F}=Z_{\kv_F}^{-1}$.
In case of an isotropic Fermi liquid one may evaluate the Fermi surface integrals by expanding $\presuper{0\!}F$ into the Legendre polynomials, 
which leads to a geometric series, i.e., Leggett's result~\cite{Legget65}.
Diagrammatic derivations of Leggett's formula were recently presented in Refs.~\cite{Wu18,Chubukov18}.
Some steps to approach the integral in Eq.~\eqref{eq:legget2} in the anisotropic case are shown in Ref.~\cite{Frigeri02}, which is however substantially more difficult.

\section{Fermi liquid relations in DMFT}\label{sec:dmft}
We apply the DMFT approximation to the Fermi liquid relations derived above and discuss simplifications.
We furthermore show how central quantities can be recovered by extrapolation from finite temperature.
\subsection{Self-consistent DMFT scheme}\label{sec:dmftapproximation}
Within the dynamical mean-field theory the Hubbard model~\eqref{eq:hubbard} is mapped to an auxiliary Anderson impurity model (AIM) with a local self-energy~\cite{Georges96}.
We denote by $g$ the (numerically exact) Green's function of the AIM and by $G$ the Green's function of the Hubbard model~\eqref{eq:hubbard} in DMFT approximation.
In our calculations we evaluate $g$ and the impurity vertex function using a modern CTQMC solver based on the ALPS libraries~\cite{ALPS2}
with improved estimators for the impurity vertex~\cite{Hafermann12,Hafermann14}.
Starting from an initial guess, the parameters of the AIM are adjusted self-consistently, until the condition,
\begin{align}
    G_{\text{loc}}=g,\label{eq:dmftsc}
\end{align}
is satisfied. Here, $G_{\text{loc}}$ indicates the local part of $G$.
The evaluation of the local component of $G$ requires an energy integration over the non-interacting density of states of the chosen lattice.
This is indeed the only dependence of the results on the original lattice. For this reason, in this work we consider a triangular lattice,
which is taken as a representative of a generic lattice where the density of states has no singularity at the Fermi level or special symmetries,
like the particle-hole symmetry of the square lattice.
We do \textit{not} consider the case of infinite dimensions because it leads to a highly specific momentum-dependence of response functions~\cite{Georges96}.

We stress that in this work we limit ourselves to the DMFT picture of a Mott transition~\cite{Georges96} which neglects non-local correlations.
The latter can lead to the opening of a correlation gap at small to intermediate interaction in the Hubbard model on the square lattice~\cite{Schaefer15,vanLoon18-2}, which will not be considered here.

\subsection{One- and two-particle self-energies}\label{sec:dmftvertices}
In DMFT the electronic self-energy is approximated with a local frequency-dependent self-energy $\Sigma_k\equiv\Sigma(\nu)$ which is obtained from the auxiliary impurity model, so that the lattice Green's function reads,
\begin{align}
    G_{\kv\nu}=&[\nu-\varepsilon_{\kv}+\mu-\Sigma(\nu)+\imath\eta]^{-1}\label{eq:gdmft}.
\end{align}
A self-consistent set of $G$ and $\Sigma$ is obtained through the self-consistent cycle described in Sec.~\ref{sec:dmftapproximation}.
Therefore, in the Fermi liquid regime the quasi-particle dispersion in Eq.~\eqref{eq:gffl} is given as
$\tilde{\varepsilon}_\kv-\mu=Z[\varepsilon_{\kv}-\mu+\Re\Sigma(0)]$,
where $Z$ is the $\kv$-independent quasi-particle weight of the DMFT approximation.

In order to evaluate the vertex function it is necessary to use an appropriate approximation to the two-particle self-energy $\Gamma$.
A consistent choice for $\Gamma$ is the functional derivative of the single-particle self-energy $\Sigma$,
$\gamma=\frac{\delta\Sigma}{\delta g}$, where $g$ is the local Green's function of the auxiliary AIM, hence,
\begin{align}
    \Gamma^{\alpha}_{kk'q}\equiv\gamma^{\alpha}_{\nu\nu'\omega}.\label{eq:gammadmft}
\end{align}

In turn, the single-particle self-energy $\Sigma$ of DMFT is given as the functional derivative of the local Luttinger Ward functional $\phi$ of the AIM, $\Sigma=\frac{\delta\phi}{\delta g}$.
In combination with the self-consistency condition~\eqref{eq:dmftsc} this is sufficient to satisfy global conservation laws at the one-particle level~\cite{Potthoff06-2}.
The choice of $\Gamma$ in Eq.~\eqref{eq:gammadmft} implies that DMFT is also conserving at the two-particle level~\cite{Baym62} and consequently satisfies the Ward identity~\cite{Hafermann14-2,Krien17}, which is a crucial element of the Fermi liquid theory
(cf. Eqs.~\eqref{eq:ward:g3statch}-\eqref{eq:lambdadynt0} and Refs.~\cite{Noziere62-1,Noziere62-2}).

Conservation laws at the two-particle level guarantee the thermodynamic consistency of approximations,
which is expressed by the Ward identities~\eqref{eq:ward:g3statch} and~\eqref{eq:ward:g3statsp}.
In DMFT we can therefore study response functions at the one-particle level (e.g., $\frac{d\langle m\rangle}{dh}$)
or at the two-particle level ($\presuper{\infty\!}X^\sz$), leading to the same result~\cite{vanLoon15} and predicting the same divergences~\cite{Janis17}.
We stress that the Ward identity is ultimately satisfied in DMFT due to the self-consistency condition~\eqref{eq:dmftsc}~\cite{Krien17}. Therefore, particular care has to be taken in the implementation, which needs to provide numerically exact convergence, which can be reasonably reached within CTQMC,
while the exact diagonalization method~\cite{Caffarel94} may lead to deviations from  thermodynamic consistency.

\subsection{Fermi liquid relations}\label{sec:flparms}
The DMFT approximation in Eqs.~\eqref{eq:gdmft} and~\eqref{eq:gammadmft} leads to several simplifications at the two-particle level.
Due to the momentum-independence of the two-particle self-energy $\gamma$, the vertex function $F$
depends only on the transferred momentum $\qv$, not on the momenta $\kv$ and $\kv'$~\footnote{
    Due to this simplification the DMFT approximation may violate exact statements about the vertex $F_{kk'q}$
    that depend on the direction of $\kv$ and $\kv'$, such as the forward scattering sum rule~\cite{Mermin67}.
}.
Therefore, the Bethe-Salpeter equation~\eqref{eq:bselimits_inv} in the limit $q\rightarrow0$ simplifies to,
\begin{align}
    \presuper{\infty\!}F^\alpha_{\nu\nu'}=\presuper{0\!}F^\alpha_{\nu\nu'}
    +\frac{1}{2\pi}\int d\nu''\presuper{0\!}F^\alpha_{\nu\nu''}R(\nu'')_\text{loc}\presuper{\infty\!}F^\alpha_{\nu''\nu'},
   \label{eq:bselimits_inv_dmft}
\end{align}
Here we have introduced the local discontinuity, $R_\text{loc}(\nu)=\frac{1}{N}\sum_\kv R_{\kv\nu}$.
Using the explicit expression for $R$ in Eq.~\eqref{eq:rdef} and for the quasi-particle dispersion $\tilde{\varepsilon}_\kv$ we may write, 
\begin{align}
    R_\text{loc}(\nu)=&-2\pi\imath Z^2\delta(\nu){D}^*(0),\label{eq:rdmft}
\end{align}
where we defined the renormalized (quasi-particle) density of states (DOS) at the Fermi level,
${D}^*(0)=\frac{1}{N}\sum_\kv\delta(\tilde{\varepsilon}_\kv-\mu)=D(0)/Z$,
and $D(0)=\frac{1}{N}\sum_\kv\delta[{\varepsilon}_\kv-\mu+\Re\Sigma(0)]$ is the interacting DOS at the Fermi level,
which coincides with the non-interacting one because of the Luttinger theorem for a momentum-independent self-energy~\cite{muellerhartmann89}.

One may now derive the usual Fermi liquid relations~\cite{Landau80,Legget65}.
Using Eq.~\eqref{eq:rdmft} we can evaluate the Bethe-Salpeter equation~\eqref{eq:bselimits_inv_dmft} at the Fermi level, $\nu=\nu'=0$,
\begin{align}
    \presuper{\infty\!}F^\alpha_{00}=\frac{\presuper{0\!}F^\alpha_{00}}{1+\mathfrak{f}^{\,\alpha}}\label{eq:f00dmft},
\end{align}
where we defined the Landau parameters as,
\begin{align}
    \mathfrak{f}^{\,\alpha}=\imath Z^2D^*(0)\presuper{0\!}F^\alpha_{00}\label{def:landauparameter},
\end{align}
which arise from the dynamic limit $\presuper{0\!}F^\alpha_{00}$ of the vertex function at the Fermi level.
Furthermore, from Eqs.~\eqref{eq:fbvertex} and~\eqref{eq:legget} we obtain,
\begin{align}
    \presuper{\infty\!}\Lambda^\alpha_{0}=\frac{1}{Z(1+\mathfrak{f}^{\,\alpha})}\label{eq:lambda0dmft},\\
    \presuper{\infty\!}X^\alpha=\frac{-2\imath D^*(0)}{1+\mathfrak{f}^{\,\alpha}}\label{eq:xdmft},
\end{align}
where we note that in DMFT the three-leg vertex $\Lambda_{kq}=\Lambda_{\nu q}$ does not depend on the momentum $\kv$, similar to the vertex function.
The Ward identity~\eqref{eq:lambdadynt0}, $\presuper{0\!}\Lambda^\alpha_{0}=Z^{-1}$, was used to derive Eq.~\eqref{eq:lambda0dmft}.

We further evaluate the double limit $q\rightarrow0$ of the response function $L_{kq}=G_kG_{k+q}\Lambda_{\nu q}$.
Note that even in DMFT $L_{kq}$ \textit{does} depend on $\kv$, due to the attached bubble.
We show in Appendix~\ref{app:lindmft} that Eq.~\eqref{eq:g3decomp} implies,
\begin{align}
    \!\presuper{\infty\!}L^{\alpha}_{k}=&\presuper{0\!}L^{\alpha}_{k}\!+\!\frac{\presuper{\infty\!}X^\alpha Z}{2}
    \!\!\left[\frac{2\pi\delta(\nu)\delta(\tilde{\varepsilon}_\kv-\mu)}{D^*(0)}\!+\!\presuper{0\!}G^2_{k}\presuper{0\!}F^\alpha_{\nu0}\right]\label{eq:g3decompdmft}\!.
\end{align}

The algebraic Fermi liquid relations~\eqref{eq:f00dmft}-\eqref{eq:xdmft} are of course well-known, however, we stress that here they arise as exact results for the DMFT
approximation to the single-band Hubbard model at zero temperature, and they are valid for any lattice dispersion.
As such, these expressions have to our best knowledge not been derived rigorously in the literature before, although they have been used~\cite{Capone02}.
We note that next to the Landau parameters $\mathfrak{f}$ of the lattice approximation one may also define \textit{impurity} Landau parameters~\cite{Mora15}.
Since the DMFT approximation is not two-particle self-consistent~\cite{Krien17}, such a quantity is in general not equivalent to $\mathfrak{f}$.

\subsection{Limit of vanishing quasi-particle weight}\label{sec:z0}
We consider the Fermi liquid relations~\eqref{eq:f00dmft}-\eqref{eq:xdmft} and equation~\eqref{eq:g3decompdmft} in the DMFT approximation
to the half-filled single-band Hubbard model~\eqref{eq:hubbard} near the interaction-driven Mott transition at zero temperature, where $Z\rightarrow0$.

First we consider the charge sector $\alpha=\ch$.
It is known that near the transition the charge susceptibility approaches zero as $\presuper{\infty\!}X^\ch\propto Z$~\cite{Kokalj13},
we can therefore deduce from the Fermi liquid formula~\eqref{eq:xdmft} that $-2\imath D^*(0)/(1+\mathfrak{f}^{\,\ch})\propto Z$.
The quasi-particle DOS $D^*(0)=Z^{-1}D(0)$ diverges at the transition $\propto Z^{-1}$,
since the bandwidth of the quasi-particle dispersion $\tilde{\varepsilon}_\kv$ shrinks to zero, and hence the symmetric Landau parameter diverges,
\begin{align}
\mathfrak{f}^{\,\ch} \propto Z^{-2}.
\end{align}

A further remarkable result follows for the forward scattering amplitude $\presuper{\infty\!}F^\alpha_{00}$ at the Fermi level.
We can express the static susceptibility in terms of $\presuper{\infty\!}F^\alpha_{00}$ by combining Eqs.~\eqref{eq:xdmft} and~\eqref{eq:f00dmft}, which leads to,
\begin{align}
    \presuper{\infty\!}X^\alpha=-2\imath{D}^*(0)\left[1-\imath Z^2{D}^*(0)\presuper{\infty\!}F^\alpha_{00}\right]\label{eq:xfromf00}.
\end{align}
Usually a divergence of the forward scattering is associated to a Pomeranchuk instability~\cite{Pomeranchuk59},
leading to the divergence of the corresponding $\presuper{\infty\!}X^\alpha$.
We see from Eq.~\eqref{eq:xfromf00} that this is indeed the case when $Z$ is finite.
However, at the Mott transition the forward scattering must diverge in order for $\presuper{\infty\!}X^\ch$ to vanish as $Z\rightarrow0$,
\begin{align}
\presuper{\infty\!}F^\ch_{00}\propto Z^{-1}.
\end{align}

In the spin channel the situation is slightly different, since $\presuper{\infty\!}X^\sz$ does not vanish at the Mott transition.
In the case that it remains finite~\footnote{
    At zero temperature the total magnetic response of the Mott insulator
    is commonly believed to be finite, due to the effective exchange $\tilde{t}^{\,2}/U$~\cite{Rozenberg94,Georges96,Hewson16}.
    However, calculations are hindered practically because the self-consistent DMFT equations are unstable in presence of a magnetic field~\cite{Hewson16}.
    A modification of the impurity problem has been suggested to circumvent this problem~\cite{Guerci19}.},
the anti-symmetric Landau parameter $\mathfrak{f}^{\,\sz}$ and the forward scattering vertex $\presuper{\infty\!}F^\sz_{00}$ both diverge $\propto Z^{-1}$.
Using the previous results one also easily evaluates equation~\eqref{eq:lambda0dmft} for the static three-leg vertex $\presuper{\infty\!}\Lambda_{0}$.
\begin{table}[t]
\begin{tabular}{ l | c | c}
    & Quantity & Scaling $Z\rightarrow0$\\\hline
    \multirow{2}{*}{Total response} & $\presuper{\infty\!}X^\ch$ & $\propto Z$\\\cline{2-3}
    & $\presuper{\infty\!}X^\sz$ & $\propto \text{const}$\\\hline
    {Landau parameter} & $\mathfrak{f}^{\,\ch}$ ($\presuper{0\!}F^\ch_{00}$) & $\propto Z^{-2}$ $(\propto Z^{-3})$\\\cline{2-3}
    (dynamic vertex)& $\mathfrak{f}^{\,\sz}$ ($\presuper{0\!}F^\sz_{00}$) & $\propto Z^{-1}$ $(\propto Z^{-2})$\\\hline
    \multirow{2}{*}{Forward scattering} & $\presuper{\infty\!}F^\ch_{00}$ & $\propto Z^{-1}$\\\cline{2-3}
    & $\presuper{\infty\!}F^\sz_{00}$ & $\propto Z^{-1}$\\\hline
    \multirow{2}{*}{Static three-leg vertex} & $\presuper{\infty\!}\Lambda^\ch_{0}$ & $\propto Z$\\\cline{2-3}
    & $\presuper{\infty\!}\Lambda^\sz_{0}$ & $\propto\text{const}$\\\hline
    {Dynamic three-leg vertex} & $\presuper{0\!}\Lambda^\ch_{0}=\presuper{0\!}\Lambda^\sz_{0}$ & $=Z^{-1}$\\\hline
    \multirow{2}{*}{Fermion-boson response} & $\presuper{\infty\!}L^\ch_{k}-\presuper{0\!}L^\ch_{k}$ & $\propto Z\times\text{fct}(k)$\\\cline{2-3}
    & $\presuper{\infty\!}L^\sz_{k}-\presuper{0\!}L^\sz_{k}$ & $\propto\text{const}\times\text{fct}(k)$\\\hline
\end{tabular}
\caption{\label{tab:scaling}
    Scaling of various quantities with the quasi-particle weight $Z\rightarrow0$
    for the DMFT approximation to the single-band Hubbard model at zero temperature.
    'const' and '$\text{fct}(k)$' [$\estimates$ function of $k=(\kv,\nu)$] are to leading order independent of $Z$.
    Results for the spin channel are valid under the assumption that the static spin susceptibility approaches a finite value as $Z\rightarrow0$.}
\end{table}

Lastly, we consider the static response function $\presuper{\infty\!}L_{k}$.
The latter is given by Eq.~\eqref{eq:g3decompdmft}, we examine the second term on its right-hand-side,
\begin{align}
    \frac{\presuper{\infty\!}X^\alpha Z}{2}
    \left[\frac{2\pi\delta(\nu)\delta(\tilde{\varepsilon}_\kv-\mu)}{D^*(0)}\!+\!\presuper{0\!}G^2_{k}\presuper{0\!}F^\alpha_{\nu0}\right].\label{eq:remainder}
\end{align}
The term in braces is proportional $Z^{-1}$, which can be seen by integrating over $\frac{1}{2\pi N}\int_{-\infty}^{+\infty} d\nu\sum_\kv$.
Therefore, in the charge channel the whole term~\eqref{eq:remainder} is proportional $\presuper{\infty\!}X^\ch\propto Z$ and thus vanishes at the Mott transition.
Using this result in Eq.~\eqref{eq:g3decompdmft} we obtain,
\begin{align}
    \presuper{\infty\!}L^\ch_{k}=\presuper{0\!}L^\ch_{k}+Z\times\text{fct}(k),\label{eq:fbresponsez0}
\end{align}
where $\text{fct}(k)$ is some function of order $\mathcal{O}(1)$ with respect to $Z$.
As a consequence, as $Z\rightarrow0$ the static charge response $\presuper{\infty\!}L^\ch_{k}$ approaches the dynamic response $\presuper{0\!}L^\ch_{k}$.
This result is of course consistent with the scaling $\presuper{\infty\!}X^\ch\propto Z$ of the charge susceptibility near the transition,
\begin{align}
    \presuper{\infty\!}X^\ch=&2\int_k\presuper{\infty\!}L^\ch_{k}=2\int_k\presuper{0\!}L^\ch_{k}+2Z\int_k\text{fct}(k)\notag\\
    =&Z\times\text{const},
\end{align}
where we used that the integral over $\presuper{0\!}L^\ch$ vanishes [see below Eq.~\eqref{eq:legget}].
A similar situation does not arise in the spin channel,
because the spin susceptibility $\presuper{\infty\!}X^\sz=2\int_k\presuper{\infty\!}L^\sz_{k}$ does not vanish at the Mott transition.

The scaling relations in the limit $Z\rightarrow0$ derived in this subsection are our main result, they are summarized in Table~\ref{tab:scaling}.

\subsection{Extrapolation from finite temperature}\label{sec:extrapolation}
The Fermi liquid relations~\eqref{eq:f00dmft}-\eqref{eq:xdmft} can be evaluated when the quasi-particle weight $Z$,
the quasi-particle DOS $D^*(0)$, and one additional quantity are known.
In our calculations this will be the total static response $\presuper{\infty\!}X$ at finite temperature.
In this subsection $\nu_n=(2n+1)\pi T$ and $\omega_m=2m\pi T$ denote fermionic and bosonic Matsubara frequencies, the labels $n,m$ will be dropped.
$k, q$ comprise momentum and Matsubara frequency, respectively.

In order to calculate $\presuper{\infty\!}X$ we use the following Bethe-Salpeter equation for the vertex function (see, e.g.,~\cite{Hafermann14-2}),
\begin{align}
    F^\alpha_{\nu\nu'}(q)=f^\alpha_{\nu\nu'\omega}+T\sum_{\nu''}f^\alpha_{\nu\nu''\omega}\tilde{X}^0_{\nu''}(q)F^\alpha_{\nu''\nu'}(q).\label{eq:bsedmftphi}
\end{align}
Here, $\tilde{X}^0_\nu(q)=\frac{1}{N}\sum_\kv (G_k-g_\nu)(G_{k+q}-g_{\nu+\omega})$ is a bubble of non-local DMFT Green's functions $G_k-g_\nu$,
where the lattice Green's function $G_k$ and the impurity Green's function $g_\nu$ are known on the Matsubara frequencies.
$f$ denotes the impurity vertex function $f$ (the impurity analogue to $F$~\footnote{
    The vertex function $f$ and the two-particle self-energy $\gamma$ of the impurity are related via the \textit{impurity}
    Bethe-Salpeter equation, $f^\alpha_{\nu\nu'\omega}=\gamma^\alpha_{\nu\nu'\omega}+T\sum_{\nu''}\gamma^\alpha_{\nu\nu''\omega}g_{\nu''}g_{\nu''+\omega}f^\alpha_{\nu''\nu'\omega}$.
}). $g$ and $f$ are known numerically exactly.

We note that in Eq.~\eqref{eq:bsedmftphi} the two-particle self-energy $\gamma$ of the impurity does not appear explicitly [cf. Appendix, Eq.~\eqref{eq:bsedmft}].
This formulation of the Bethe-Salpeter equation is reminiscent of the dual fermion and dual boson approaches~\cite{Rubtsov08,Rubtsov12},
we use it here because $\gamma$ may be divergent in the non-critical Fermi liquid regime~\cite{Gunnarsson17},
which is to our best knowledge not the case for $f$.

After $F$ has been calculated, we obtain the three-leg vertex as,
\begin{align}
    \Lambda^\alpha_{\nu}(q) = 1 + \frac{T}{N}\sum_{\kv'\nu'}F^\alpha_{\nu\nu'}(q)G_{k'}G_{k'+q},\label{eq:lambda_dmft}
\end{align}
and the response function $L^{\alpha}_{k}(q)=G_kG_{k+q}\Lambda^\alpha_{\nu}(q)$, both given at the Matsubara frequencies.
Finally, the total response is given by $\presuper{\infty\!}X^\alpha=2\imath\frac{T}{N}\sum_{k}L^{\alpha}_{k}(q=0)$.

Note that the limit $q\rightarrow0$ is not ambiguous on the Matsubara frequencies, since they are discrete,
and it always leads to the static homogeneous limit $\mathfrak{r}=\infty$.
In order to evaluate dynamic limits we consider the finite frequencies $\omega_1=2\pi T$ and $\nu_{-1}=-\pi T\equiv\bar{\nu}$
[this notation will be used throughout] at low temperature.
From these frequencies we can obtain the dynamic three-leg vertex $\presuper{0\!}\Lambda_0$ at the Fermi level in the limit $T\rightarrow0$.
To see this, we use the Ward identity for the Matsubara three-leg vertex $\Lambda$, which is derived in Appendix~\ref{app:lambdadyn}.
Evaluating the latter at $\bar{\nu}$ and $\omega_1$ yields
\begin{align}
    \Lambda^\alpha_{\bar{\nu}}(\qv_0=\mathbf{0},\omega_1)=&1-\frac{\Sigma_{-\bar{\nu}}-\Sigma_{\bar{\nu}}}{-2\imath\bar{\nu}}=1-\frac{\Im\Sigma(\pi T)}{\pi T},\label{eq:lambdaz}
\end{align}
where we used $\omega_1=-2\bar{\nu}$ and $\Sigma(-\bar{\nu})=\Sigma^*(\bar{\nu})$.
The right-hand-side of Eq.~\eqref{eq:lambdaz} approaches $Z^{-1}$ in the limit $T\rightarrow0$~\cite{Serene91,Arsenault12},
which recovers the zero temperature result in Eq.~\eqref{eq:lambdadynt0}.

In fact, we show in Appendix~\ref{app:ac:ward} that the Ward identity can be used to perform the analytical continuation of the three-leg vertex $\Lambda$,
or of the respective response function $L$, from Matsubara frequencies $\nu_n$ and $\omega_m$ to any pair of real frequencies $\nu$ and $\omega$.

In our numerical results we use $Z^{-1}\approx1-\frac{\Im\Sigma(\pi T)}{\pi T}$ at finite temperature as an approximation.
Similarly, we approximate the density of states at the Fermi level as~\cite{vanLoon14},
\begin{align}
    D(0)\approx-(\pi T)^{-1}g\left(\tau=1/2T\right),\label{eq:dosatfermi}
\end{align}
where $g$ is the impurity Green's function and $\tau$ the imaginary time. The quasi-particle density of states is then obtained as $D^*(0)=Z^{-1}D(0)$.
We note that these approximations become exact in the limit $T\rightarrow0$. We refer to $D(0)$ also as $\text{DOS}(0)$.

\section{The Mott transition on the triangular lattice}\label{sec:mott}
In Sec.~\ref{sec:z0} we obtained the scaling relations for the limit $Z\rightarrow0$ comprised in Table~\ref{tab:scaling}.
In the following we verify several of these results in DMFT calculations for the half-filled Hubbard model~\eqref{eq:hubbard} on the triangular lattice.
We stress that while our DMFT results were obtained at finite temperature,
our main aim is to make conclusions about the Mott transition in the limit $T\rightarrow0$.

In this section, unless clearly marked differently, we consider Matsubara correlation functions and vertices, $G^m, X^{m,\alpha}, ...$, and so on.
The label $m$ will be dropped in the following.
\begin{figure}
    \begin{center}
    \includegraphics[width=0.49\textwidth]{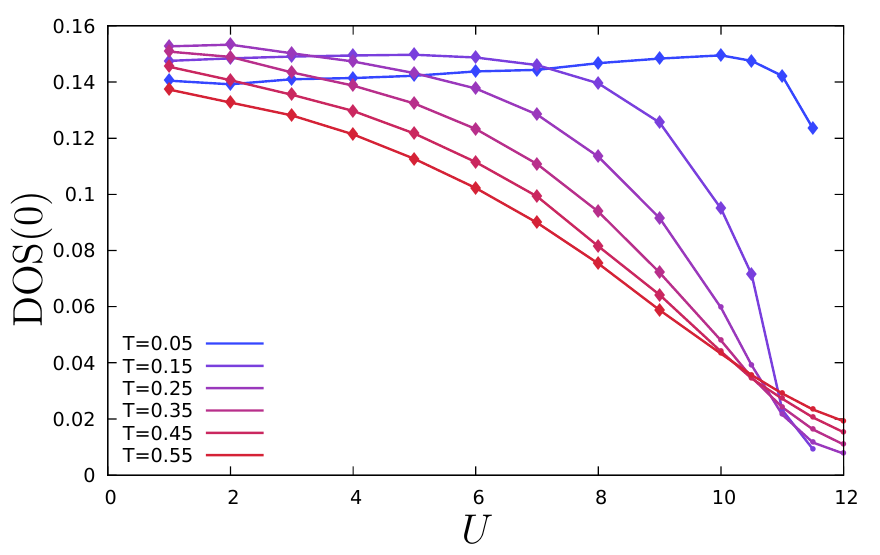}
    \end{center}
    \vspace{-.3cm}
    \caption{\label{fig:gtau} (Color online) Approximate DOS of the half-filled Hubbard model~\eqref{eq:hubbard} on the triangular lattice at the Fermi level.
%    The inflection point of each curve indicates the value $U_M(T)$ of the Mott crossover/transition.
    Diamonds and small dots indicate metallic and insulating solutions, respectively.
    This labeling was obtained from Fig.~\ref{fig:ev_dyn}, as described in Sec.~\ref{sec:eigenvalues},
    and is used consistently also in Figs.~\ref{fig:suscs},~\ref{fig:rlat}, and~\ref{fig:fdyn_ch}.
    Black-white version: In the center $T$ increases from right to left.
    }
\end{figure}

\subsection{Spectral weight at the Fermi level and static susceptibility}
To set the stage, we firstly identify the metal and Mott regimes of the Hubbard model~\eqref{eq:hubbard} within the DMFT approximation.
We note that near the Mott transition/crossover solutions of smaller $U$ were used as an input for the DMFT loop at larger $U$.
We do not consider the coexistence of insulating and metallic solutions or the first order critical line at low temperature,
see, for example, Refs.~\cite{Werner07,Balzer09}.
\begin{figure}
    \begin{center}
    \includegraphics[width=0.49\textwidth]{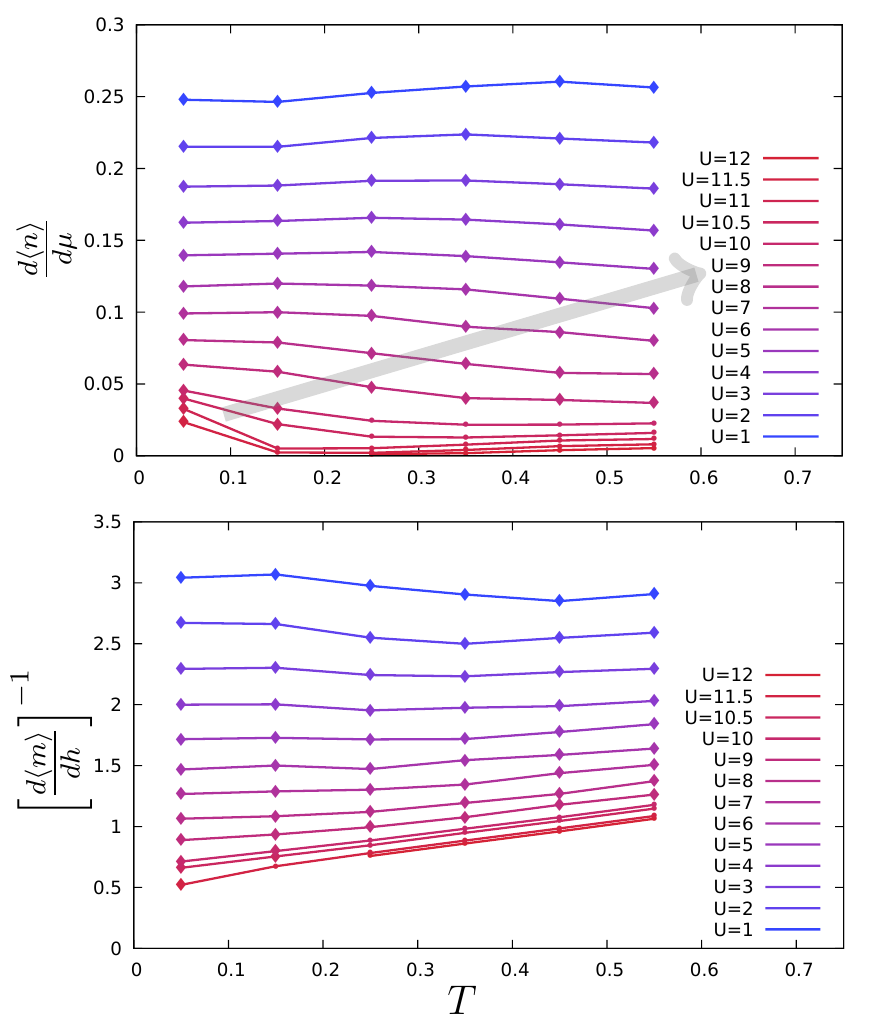}
    \end{center}
    \vspace{-.3cm}
    \caption{\label{fig:suscs} (Color online) Top: Charge susceptibility $\frac{d\langle n\rangle}{d\mu}$ as a function of the temperature $T$.
    The arrow indicates the evolution of the coherence temperature $T_\text{coh}$ of the Fermi liquid with $U$.
    Bottom: Inverse of the spin susceptibility $\frac{d\langle m\rangle}{dh}$.
    Labels as in Fig.~\ref{fig:gtau}. Black-white version: In both panels $U$ increases from top to bottom.
    }
\end{figure}

Figs.~\ref{fig:gtau} and~\ref{fig:suscs} show the spectral weight at the Fermi level and the susceptibilities $\presuper{\infty\!}{X}^\alpha$, respectively.
In both figures diamonds are used to label metallic solutions, whereas small circles are used for insulating ones.
The labeling was obtained using a criterion that is introduced in Sec.~\ref{sec:eigenvalues}.

We begin with Fig.~\ref{fig:gtau}, which shows the approximate spectral weight at the Fermi level $D(0)=-g(\tau=1/2T)/(\pi T)$~\footnote{
For $T>0$ the interacting density of states $D(0)$ of the Fermi liquid is in general very different from the non-interacting one,
since the pinning of its value to the non-interacting DOS according to the Luttinger theorem is realized only at very low $T$.}
as a function of $U$ for temperatures $0.05\leq T\leq 0.55$ in units of the hopping $\tilde{t}=1$. $g$ is the impurity Green's function.
The lines show inflection points at elevated $U$, which indicate the interaction $U_M(T)$ of the Mott crossover/transition.
Below $U_M(T)$ Fig.~\ref{fig:gtau} shows for lower temperature that the spectral weight at the Fermi level increases with $U$.
This is a particularity of the triangular lattice, whose quasi-particle peak and van Hove singularity merge near the critical interaction~\cite{Aryanpour06}.
In the limit $T\rightarrow0$ the spectral weight at the Fermi level vanishes completely for $U\geq U_M(T=0)\approx 12$~\cite{Aryanpour06}.

The top panel of Fig.~\ref{fig:suscs} shows the static homogeneous charge susceptibility,
$\frac{d\langle n\rangle}{d\mu}=\imath\presuper{\infty\!}X^\ch$,
where we calculated $\presuper{\infty\!}X^\ch$ at finite temperature according to Sec.~\ref{sec:extrapolation}
[the factor $\imath$ accommodates to the zero temperature definition~\eqref{eq:suscs}].
Data points in the Mott regime tend toward zero with decreasing temperature, whereas those in the metallic regime tend towards finite values.
For moderate interaction $\frac{d\langle n\rangle}{d\mu}$ shows an upturn as $T$ is lowered,
which occurs near the coherence temperature of the Fermi liquid~\cite{Mezio17}.
This pattern crosses the panel diagonally from the bottom left to the top right (arrow), near $U_M(T)$ it leads to a $T$-driven insulator-to-metal crossover.
Above $U_M(T=0)$ a re-entry into the Fermi liquid at low temperature does no longer occur.
At $T=0$ the charge susceptibility then vanishes exactly~\cite{Werner07}.

The bottom panel of Fig.~\ref{fig:suscs} shows the inverse of the spin susceptibility, $\frac{d\langle m\rangle}{dh}=\imath\presuper{\infty\!}X^\sz$.
The latter does not vanish in the Mott phase and it appears to be unaffected by the transition, in agreement with early DMFT results~\cite{Rozenberg94}.
This can be seen well for $U=10, 10.5$, and $11$, where $\frac{d\langle m\rangle}{dh}$ changes continuously at the $T$-driven crossover.
\begin{figure}
    \begin{center}
    \includegraphics[width=0.49\textwidth]{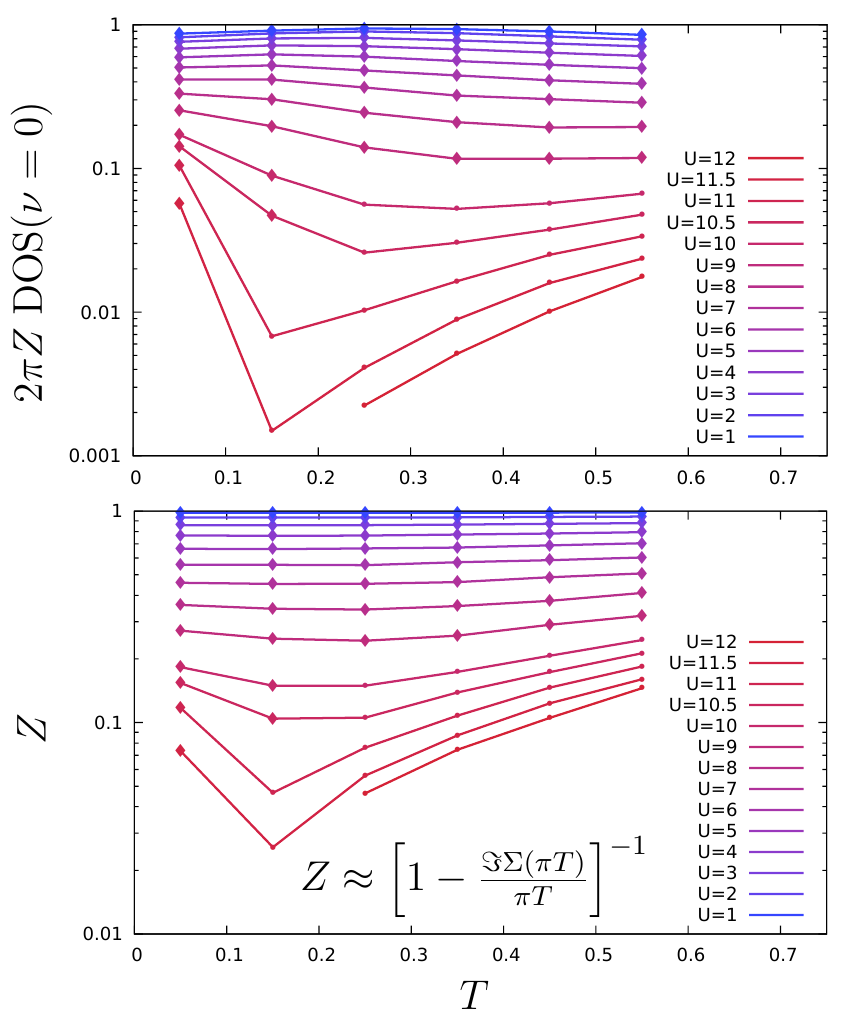}
    \end{center}
    \vspace{-.3cm}
    \caption{\label{fig:rlat} (Color online)
    Top: Absolute value of the discontinuity $R_\text{loc}(\nu=0)=-2\pi\imath Z \,\text{DOS}(\nu=0)\delta(0)$ [cf. Eq.~\eqref{eq:rdmft}]
    in DMFT calculations at finite temperature. At $T=0$ the discontinuity is finite in the metal and zero in the Mott insulator.
    Bottom: Approximation to the quasi-particle weight $Z$. Labels as in Fig.~\ref{fig:gtau}.
    Black-white version: In both panels $U$ increases from top to bottom.
     }
\end{figure}

\subsection{Discontinuity and quasi-particle weight}\label{sec:rplus}
We show in the top panel of Fig.~\ref{fig:rlat} the singular value $R_\text{loc}(\nu=0)=-2\pi\imath Z \,\text{DOS}(\nu=0)\delta(0)$ of the local discontinuity [cf. Eq.~\eqref{eq:rdmft}].
To calculate this quantity at finite temperature we use Eq.~\eqref{eq:dosatfermi}, as before,
and $Z^{-1}\approx 1-\frac{\Im\Sigma(\pi T)}{\pi T}$ [cf. Eq.~\eqref{eq:lambdaz}].
Similar to the charge susceptibility in Fig.~\ref{fig:suscs}, Fig.~\ref{fig:rlat} shows the re-entry into the Fermi liquid at low temperature for $U<U_M(T=0)$.
For insulating solutions the behavior of $R_\text{loc}(\nu=0)$ is consistent with its vanishing at $T=0$ for $U>U_M(T=0)$.

The bottom panel shows the approximation for $Z$ from the first Matsubara energy.
The results of this work [in particular Fig.~\ref{fig:scaling}] do not change qualitatively when $Z$
is determined via polynomial extrapolation of the self-energy $\Sigma$ to the Fermi level.

\subsection{Divergence of the dynamic three-leg vertex}\label{sec:lambdadiv}
We verify the divergence of the dynamic limit of the three-leg vertex $\presuper{0\!}\Lambda=Z^{-1}$ at the Mott transition.
Since this is an analytical statement, it is certain that this divergence occurs as $Z\rightarrow0$.
We show in the following that this is a direct consequence of the Ward identity.

The DMFT self-consistency condition~\eqref{eq:dmftsc} leads to the equivalence of the Ward identity
of the Hubbard model~\eqref{eq:hubbard} with the Ward identity of the Anderson impurity model,
which is a local relation~\cite{Hafermann14-2,Krien17},
\begin{align}
    \Sigma_{\nu+\omega}-\Sigma_{\nu}=T\sum_{\nu'}\gamma^\alpha_{\nu\nu'\omega}[g_{\nu'+\omega}-g_{\nu'}].\label{eq:localward}
\end{align}
Here, $\gamma^\alpha$ is the two-particle self-energy of the AIM, $g$ is the impurity Green's function.
We note that both $\gamma^\ch$ and $\gamma^\sz$ satisfy this equation.

\begin{figure}
    \begin{center}
    \begin{tikzpicture}
        \node[anchor=south west,inner sep=0] (image) at (0,0) {\includegraphics[width=0.49\textwidth]{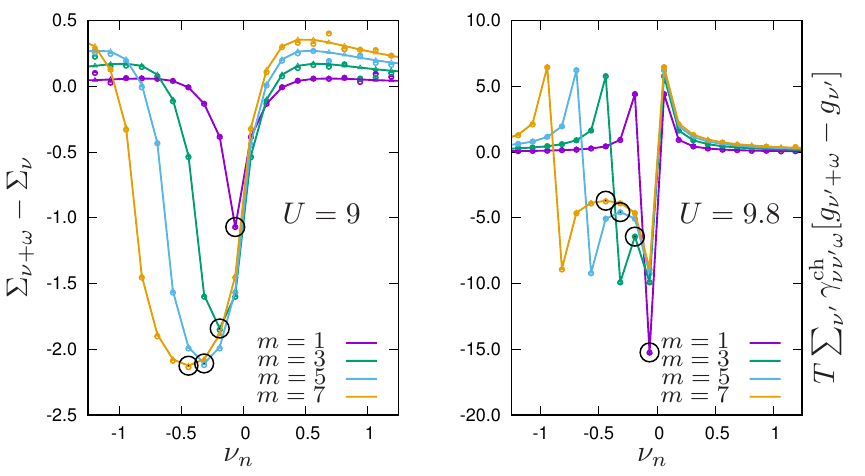}};
        \begin{scope}[x={(image.south east)},y={(image.north west)}]
            \node[thick] at (0.5,1.05) {(square lattice, Figure reprinted from Ref.~\cite{Hafermann14-2})};
        \end{scope}
    \end{tikzpicture}
    \end{center}
    \vspace{-.3cm}
    \caption{\label{fig:localward} (Color online) Imaginary part of the Ward identity~\eqref{eq:localward} in the metal (left) and in the Mott insulator (right)
    for several fixed values of the bosonic frequency $\omega_m=2m\pi T$.
    Discrepancies between the left-hand-side (lines) and right-hand-side (symbols) are due to numerical noise.
    Black circles mark data points at $\nu_n=-\omega_m/2$, where $\Sigma(\nu_n+\omega_m)-\Sigma({\nu_n})=-2\imath\Im\Sigma(\nu_n)$,
    which diverges in the Mott phase as $T\rightarrow0$.
    }
\end{figure}

Fig.~\ref{fig:localward} shows a numerical validation of Eq.~\eqref{eq:localward} in the metal (left panel) and in the insulator (right panel).
Note that the figure corresponds to a DMFT calculation for the square lattice from Ref.~\cite{Hafermann14-2} at $T=0.08$ in our units of the hopping ($\tilde{t}=1$).

In order to demonstrate the significance of the Ward identity for the divergence of $\presuper{0\!}\Lambda$ at the Mott transition
we have marked in Fig.~\ref{fig:localward} those combinations of the Matsubara frequencies $\nu$ and $\omega$ with black circles that satisfy the constraint $\omega=-2\nu$.
Evaluating the left-hand-side of Eq.~\eqref{eq:localward} at these points simply yields $-2\imath\Im\Sigma(\nu)$.
The slope of this function near the Fermi level changes its sign at the Mott transition/crossover, which therefore directly manifests in the Ward identity,
in fact, in the limit $T\rightarrow0$ as a divergence of $\Im\Sigma(\pi T)$.

However, the Ward identity~\eqref{eq:localward} is a relation between the one- and two-particle self-energies $\Sigma$ and $\gamma$.
We can therefore expect to find divergences at the Mott transition and within the Mott phase also at the two-particle level.
Indeed, we show in Appendix~\ref{app:lambdadyn} that Eq.~\eqref{eq:localward} directly implies,
\begin{align}
    \Lambda^\alpha_{\nu}(\qv_0,\omega)=&1-\frac{\Sigma_{\nu+\omega}-\Sigma_{\nu}}{\imath\omega},\;\;\omega\neq0.\label{eq:lambdaward}
\end{align}
As discussed in Sec.~\ref{sec:extrapolation}, we can evaluate this relation at $\bar{\nu}=-\pi T$ and $\omega_1=2\pi T$ and take the limit $T\rightarrow0$ to recover
the dynamic limit of the causal three-leg vertex at the Fermi level,
\begin{align}
    \Lambda^{m,\alpha}_{\bar{\nu}}(\qv_0,\omega_1)=\presuper{0\!}\Lambda^{c,\alpha}_{\nu=0}=&\frac{1}{Z}, \;\;\;T\rightarrow0.
\end{align}
Here, the labels $m$ and $c$ indicate the Matsubara or the causal three-leg vertex, respectively.
These should not be confused, since in general an analytical continuation is required to recover the causal vertex $\Lambda^c$
from the Matsubara vertex $\Lambda^m$ [see Appendix~\ref{app:ac:ac}].

As a consequence of the Ward identity~\eqref{eq:localward} DMFT captures the divergence of $\presuper{0\!}\Lambda^{c}$
at the critical interaction $U_M(T=0)$ of the Mott transition. The divergence occurs both in the charge and in the spin channel.

\subsection{Landau parameters}\label{sec:lparm}
\begin{figure}
    \begin{center}
    \includegraphics[width=0.49\textwidth]{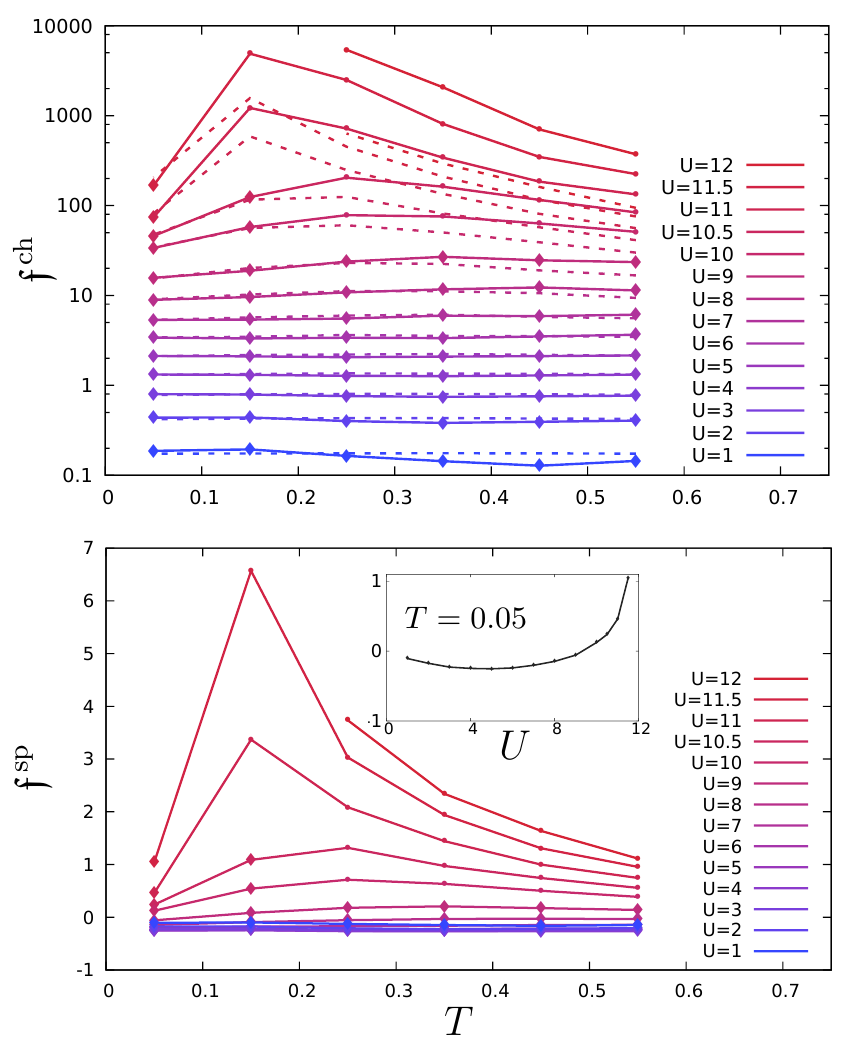}
    \end{center}
    \vspace{-.3cm}
    \caption{\label{fig:fdyn_ch} (Color online) Top: The symmetric Landau parameter $\mathfrak{}f^{\,\ch}$ calculated from Eq.~\eqref{eq:ffromx} (bold lines) as a function of temperature.
    The dashed lines indicate $Z^2 D^*(0)F^\ch_{\bar{\nu}\bar{\nu}}(\qv_0,\omega_1)$ where $\bar{\nu}=-\pi T$ and $\omega_1=2\pi T$,
    which shows agreement with $\mathfrak{f}^{\,\ch}$ at small $T$.
    Bottom: The anti-symmetric Landau parameter $\mathfrak{f}^{\,\sz}$. Labels as in Fig.~\ref{fig:gtau}.
    The inset shows $\mathfrak{f}^{\,\sz}$ as a function of $U$ for $T=0.05$.
    Black-white version: In the top panel $U$ increases from bottom to top, for bottom panel cf. inset.
    }
\end{figure}
\begin{figure}
    \begin{center}
    \includegraphics[width=0.49\textwidth]{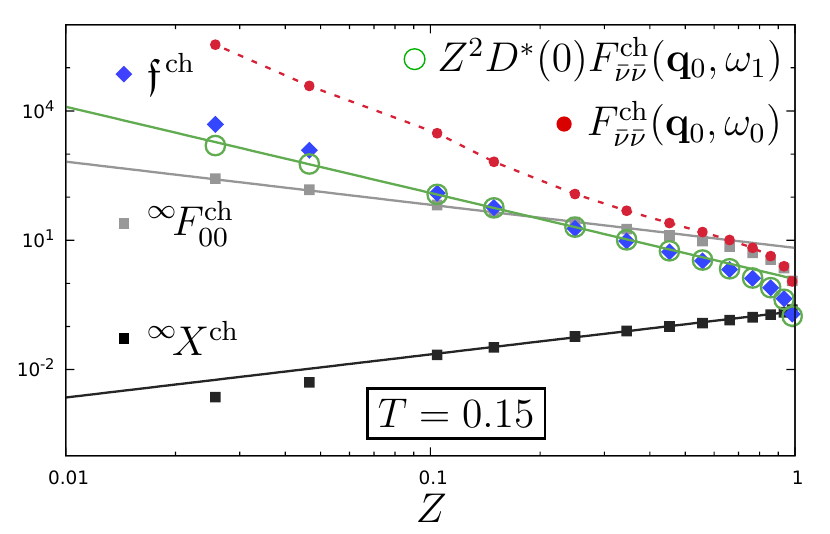}
    \end{center}
    \vspace{-.3cm}
    \caption{\label{fig:scaling} (Color online) Scaling of the symmetric Landau parameter with the quasi-particle weight $Z$ at $T=0.15$ (blue diamonds).
     Open green circles show the quantity $Z^2D^*(0)\Re F^{\ch}_{\bar{\nu}\bar{\nu}}(\qv_0,\omega_1)$ [cf. Fig.~\ref{fig:fdyn_ch}],
     green line shows fit of its \textit{inverse} with $aZ^{2}$, which confirms $\mathfrak{f}^{\,\ch}\propto Z^{-2}$.
     Black squares show the static charge susceptibility, black line a fit with $aZ$. % yields $b\approx1.22(8)$ (black line).
     Gray squares show the forward scattering vertex \mbox{$\imath\protect\presuper{\infty\!}F^\ch_{00}$} calculated from Eq.~\eqref{eq:f00fromx},
     gray line shows fit of its \textit{inverse} with $aZ$. Fits were performed for $0\leq Z\leq0.3$.
     Full red circles show the static Matsubara vertex $\Re F^\ch_{\bar{\nu}\bar{\nu}}(\qv_0,\omega_0=0)$.
    }
\end{figure}
We evaluate the Landau parameters $\mathfrak{f}$ defined in Eq.~\eqref{def:landauparameter} using Eq.~\eqref{eq:xdmft},
which allows to calculate $\mathfrak{f}$ from the quasi-particle weight $Z$, from the quasi-particle DOS $D^*(0)=Z^{-1}D(0)$,
where $D(0)$ is the \textit{non-interacting} DOS, and from the total response $\presuper{\infty\!}X^\alpha$ as,
\begin{align}
    \mathfrak{f}^{\,\alpha}=-1-2\imath D^*(0)/\presuper{\infty\!}X^\alpha.\label{eq:ffromx}
\end{align}
Bold lines in the top panel of Fig.~\ref{fig:fdyn_ch} show the symmetric Landau parameter $\mathfrak{f}^{\,\ch}$,
which grows rapidly and monotonously with the interaction $U$.
As a function of the temperature $\mathfrak{f}^{\,\ch}$ extrapolates towards finite values in the Fermi liquid (diamonds),
insulating solutions (small dots) are consistent with its divergence at $T=0$ above $U_M(T=0)$.
For comparison dashed lines in Fig.~\ref{fig:fdyn_ch} also show $Z^2 D^*(0)\Re F^\ch_{\bar{\nu}\bar{\nu}}(\qv_0,\omega_1)$,
where $F^\ch$ is the Matsubara vertex function at $\qv_0=\mathbf{0}$ and at the \textit{finite} bosonic frequency $\omega_1=2\pi T$.
This quantity shows a remarkable agreement with $\mathfrak{f}^{\,\ch}=\imath Z^2 D^*(0)\presuper{0\!}F^\ch_{00}$ [cf. Eq.~\eqref{def:landauparameter}]
at low temperature for all interactions, as the finite Matsubara frequency $\bar{\nu}=-\pi T$ approaches the Fermi level.
This agreement is non-trivial, since in general an analytical continuation is required to recover the dynamic vertex $\presuper{0\!}F^\ch_{00}$ from the Matsubara frequencies.
However, apparently these quantities are directly related in the limit $T\rightarrow0$. 

The bottom panel of Fig.~\ref{fig:fdyn_ch} shows the anti-symmetric Landau parameter $\mathfrak{f}^{\,\sz}$.
The latter remains small compared to $\mathfrak{f}^{\,\ch}$ and has a non-monotonous dependence on $U$ and $T$.
For insulating solutions (small dots) the computed $\mathfrak{f}^{\,\sz}$ is consistent with a divergence at $T=0$,
however, we could not reach low enough temperatures to confirm this trend over several orders of magnitude, see also Sec.~\ref{sec:eigenvalues}.
We are therefore not able to confirm the expected divergence of $\mathfrak{f}^{\,\sz}$.

The expression for the Landau parameters in Eq.~\eqref{eq:ffromx} is rigorous only at $T=0$.
Within the temperature range of our calculations a quantitative analysis of $\mathfrak{f}$ is therefore only reliable at small to moderate $U$,
where its dependence on the temperature is weak enough for an extrapolation to $T=0$ [see Fig.~\ref{fig:fdyn_ch}].

However, our data allow to make several qualitative statements:
(i) Both $\mathfrak{f}^{\,\ch}$ and $\mathfrak{f}^{\,\sz}$ are strictly larger than $-1$, which means that Pomeranchuk instabilities do not occur, as expected.
(ii) The trend to a divergence at $T=0$ at the Mott transition is much stronger in the symmetric Landau parameter $\mathfrak{f}^{\,\ch}$
than in the anti-symmetric $\mathfrak{f}^{\,\sz}$. This is consistent with the discussion in Sec.~\ref{sec:z0},
which implies a scaling of these quantities with the quasi-particle weight $\propto Z^{-2}$ and $\propto Z^{-1}$, respectively.
(iii) Fig.~\ref{fig:scaling} shows that at $T=0.15$ the symmetric Landau parameter $\mathfrak{f}^{\,\ch}$ indeed roughly scales $\propto Z^{-2}$.
The figure also shows $Z^2D^*(0)F^\ch_{\bar{\nu}\bar{\nu}}(\qv_0,\omega_1)$, which is in good agreement with $\mathfrak{f}^{\,\ch}$
according to Fig.~\ref{fig:fdyn_ch}, and which confirms the $\propto Z^{-2}$ scaling accurately (see green line).

We also discuss the divergence of the static charge vertex function that was predicted in Sec.~\ref{sec:z0}.
To this end, we solve Eq.~\eqref{eq:xfromf00} for $\presuper{\infty\!}F^\ch_{00}$,
\begin{align}
    \imath\presuper{\infty\!}F^\ch_{00}=\frac{1}{Z^2 D^*(0)}+\frac{\presuper{\infty\!}X^\ch}{2\imath[ZD^*(0)]^2}.\label{eq:f00fromx}
\end{align}
This quantity is marked in Fig.~\ref{fig:scaling} with gray squares and scales $\propto Z^{-1}$,
whereas black squares indicate the total charge response $\presuper{\infty\!}X^\ch$, which indeed vanishes simultaneously $\propto Z$.
Red circles in Fig.~\ref{fig:scaling} also mark our result for the static Matsubara vertex $\Re F^\ch_{\bar{\nu}\bar{\nu}}(\qv_0,\omega_0)$ for $\bar{\nu}=-\pi T$,
which shows a scaling of roughly $\propto Z^{-3}$, whereas one may expect agreement with the scaling $\propto Z^{-1}$ of $\presuper{\infty\!}F^\ch_{00}$.
There are several possible explanations for the mismatch:

Firstly, it may arise due to a subtlety in the analytical continuation of the vertex function.
To perform the latter, $F$ has to be considered within up to $8$ separate analytical regions of the $\mathbb{C}^3$-space
spanned by its three frequency indices~\cite{Oguri01,Eliashberg61} [see also Appendix~\ref{app:ac:ac}].
It can be expected that the value of $\presuper{\infty\!}F^\ch_{00}$ at the Fermi level is recovered at low temperature
as a combination of several Matrix elements $F^\ch_{\nu\nu'}(\qv_0,\omega_0)$ of the Matsubara vertex, for example, $\nu=\pm\pi T, \nu'=\pm\pi T$.
Therefore, the cancellation of a $\propto Z^{-3}$ dependence of $F$ may occur.

Secondly, among the divergences that are indicated in Fig.~\ref{fig:scaling} the one of $F^\ch_{\bar{\nu}\bar{\nu}}(\qv_0,\omega_0)$ was the most difficult to verify.
In our CTQMC calculations at low temperature the deviation of the density $\langle n\rangle$ from half filling had to be less than $10^{-6}$,
otherwise this quantity often showed a sign change, it is therefore subjected to a large error.

Thirdly, the scaling of $F^\ch_{\bar{\nu}\bar{\nu}}(\qv_0,\omega_0)$ may be sensitive to the finite temperature.
The coherence temperature $T_{\text{coh}}$ of the Fermi liquid indeed becomes very small near the Mott regime,
as indicated by the arrow in the top panel of Fig.~\ref{fig:suscs}.
Furthermore, due to the finite temperature a scaling analysis was not possible in the spin channel,
since the divergences of $\mathfrak{f}^{\,\sz}$ and $\presuper{\infty\!}F^\sz_{00}$
in Table~\ref{tab:scaling} for $T=0$ are apparently visible only at very low temperature $T\lesssim0.05$, see also the following Sec.~\ref{sec:eigenvalues}.

\subsection{Character of the divergent scatterings}\label{sec:eigenvalues}

We have seen in Sec.~\ref{sec:lambdadiv} that a divergence of the dynamic three-leg vertex $\presuper{0\!}\Lambda$ occurs as $Z\rightarrow0$.
In fact, this divergence can only occur when also the dynamic vertex function $\presuper{0\!}F$ diverges, since the latter gives rise to $\presuper{0\!}\Lambda$
by attaching a bubble [cf. Fig.~\ref{fig:3leg} a)], which is finite.
Here we consider the leading eigenvalue of the Bethe-Salpeter equation~\eqref{eq:bsedmftphi},
which was used to calculate the vertex function $F^\alpha_{\nu\nu'}(\qv_0,\omega)$.
This will reveal the driving factors behind its divergence at finite bosonic frequency, we consider $\omega=\omega_1=2\pi T$.

\begin{figure}
    \begin{center}
    \includegraphics[width=0.49\textwidth]{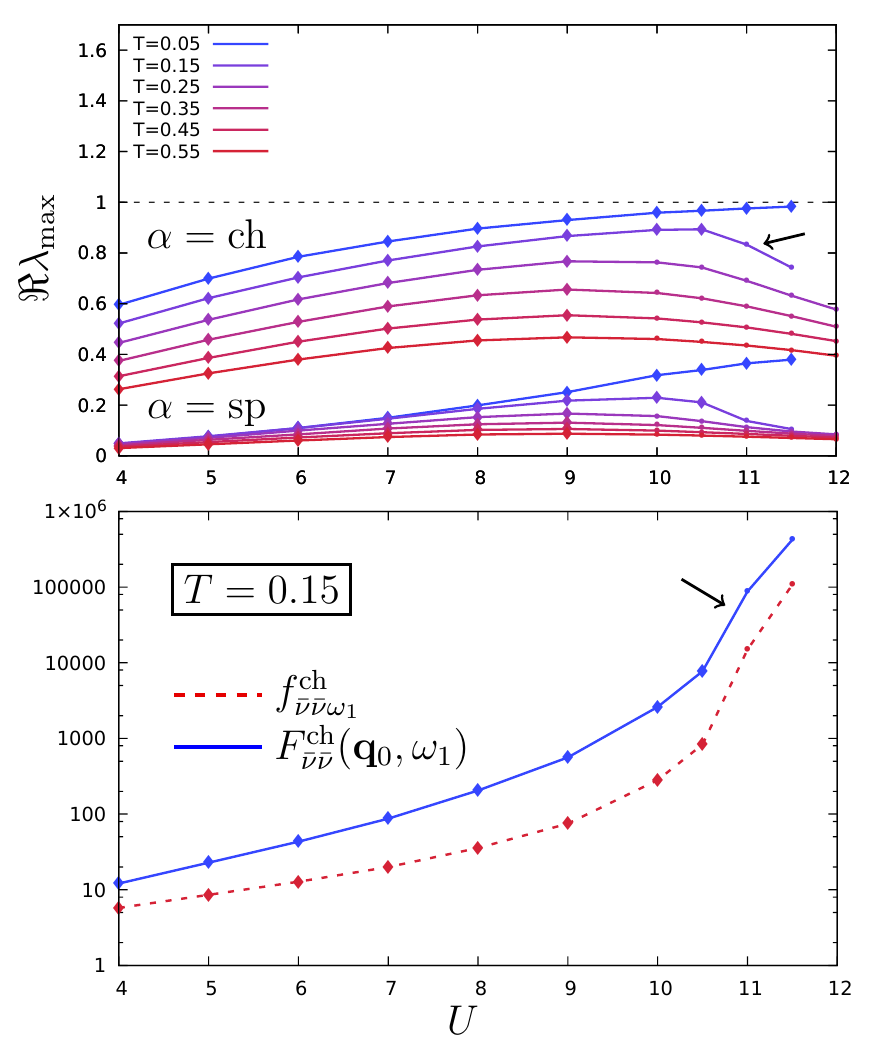}
    \end{center}
    \vspace{-.3cm}
    \caption{\label{fig:ev_dyn} (Color online) Top: Leading eigenvalue $\Re\lambda_\text{max}$ of the Bethe-Salpeter equation~\eqref{eq:bsedmftphi} for $q=(\qv_0,\omega_1)$
    in the charge (upper data set) and spin (lower data set) channel.
    The maximum $d\Re\lambda_\text{max}(U)/dU=0$ of the charge channel distinguishes metallic (diamonds) and insulating solutions (small dots)
    in Figs.~\ref{fig:gtau},~\ref{fig:suscs},~\ref{fig:rlat}, and~\ref{fig:fdyn_ch}.
    Bottom: The charge vertex $F^\ch_{\bar{\nu}\bar{\nu}}(\qv_0,\omega_1)$ (blue) and the impurity vertex $f^\ch_{\bar{\nu}\bar{\nu}\omega_1}$ (dashed red) for $T=0.15$.
    Notice the steep increase beyond $U_M(T=0.15)\gtrsim10.5$, while $\Re\lambda_\text{max}$ in the upper panel drops (arrows).
    Black-white version: In top panel $T$ increases from top to bottom.
    }
\end{figure}

The Bethe-Salpeter equation~\eqref{eq:bsedmftphi} represents the repeated application of the $\nu,\nu'$-matrix
$A^\alpha_{\nu\nu'}(\qv,\omega)=Tf^\alpha_{\nu\nu'\omega}\tilde{X}^0_{\nu'}(\qv,\omega)$ upon itself.
Here, $f$ is the impurity vertex function and $\tilde{X}^0$ is the non-local bubble defined below Eq.~\eqref{eq:bsedmftphi}.
A divergence of the lattice vertex function $F$ may occur for two reasons: (i) The leading eigenvalue of the matrix $A$ approaches unity.
(ii) The impurity vertex function $f$ diverges.

The top panel of Fig.~\ref{fig:ev_dyn} shows the leading eigenvalue $\Re\lambda_\text{max}$ of $A^\alpha(\qv_0,\omega_1)$ as a function of the interaction $U$.
The upper set of lines belongs to the charge channel $\alpha=\ch$, the lower set to the spin channel $\alpha=\sz$.
For each temperature $T$ the curve $\Re\lambda_\text{max}(U)$ has a clearly defined maximum that lies at smaller $U$ for larger $T$.
We will argue in the following that this maximum lies at the critical interaction $U_M(T)$ of the Mott transition/crossover.

Let us consider first that we approach the Mott transition from the Fermi liquid side at $T=0$.
On this side the divergence of $F(\qv_0,\omega_1)$ must be caused due to $\Re\lambda_\text{max}\rightarrow1$,
since the building blocks $f$ and $\tilde{X}^0$ of the Bethe-Salpeter equation are finite in the Fermi liquid.
The top panel of Fig.~\ref{fig:ev_dyn} shows that for $T=0.05$ the leading eigenvalue is indeed very close to unity as $U\rightarrow U_M(T=0.05)\lesssim 12$.

\begin{figure}
    \begin{center}
    \includegraphics[width=0.49\textwidth]{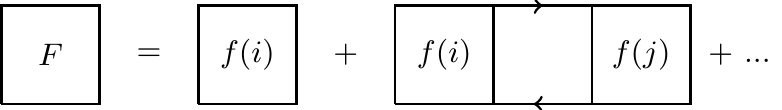}
    \end{center}
    \vspace{-.3cm}
    \caption{\label{fig:bse_realspace} The expanded Bethe-Salpeter equation~\eqref{eq:bsedmftphi} in real space,
    it connects the vertices $f(i)$ via the non-local DMFT Green's functions $G_{ij}-g\delta_{ij}$.
    As $U\rightarrow U_M$ at zero temperature, the entire sum on the right-hand-side is divergent,
    whereas for $U>U_M$ each vertex $f$ diverges.
    }
    \end{figure}

This shows that on the Fermi liquid side of the transition the driving force behind the divergence is a series
of many scattering events at different lattice sites.
This can be understood considering the Bethe-Salpeter equation in real space,
where it connects the local vertices $f(i)$ and $f(j)$ at lattice sites $i$ and $j$ via the non-local DMFT Green's function $G_{ij}-g\delta_{ij}$.
This is shown in Fig.~\ref{fig:bse_realspace}.

Let us now consider that we enter the Mott phase. Within this phase the dynamic vertices must remain divergent due to $\presuper{0\!}\Lambda=Z^{-1}$,
since $Z$ is zero throughout the insulator.
The question is therefore which mechanism sustains the divergence for $U>U_M(T=0)$.
Mathematically this could be achieved if $\Re\lambda_\text{max}$ was exactly unity everywhere in the insulator.
However, our DMFT results in Fig.~\ref{fig:ev_dyn} at finite temperature suggest that the leading eigenvalue $\Re\lambda_\text{max}(U)$ \textit{decreases} beyond $U_M(T)$.

We therefore propose a scenario for $T=0$ where $\Re\lambda_\text{max}=1$ is only realized exactly at $U=U_M(T=0)$.
Beyond this point the Bethe-Salpeter equation diverges no longer due to scattering events at different lattice sites but because
each of its building blocks $f$ diverges for $U>U_M(T=0)$.
This scenario seems likely because we find at finite temperature that the drop of the leading eigenvalue at $U_M(T)$
does not lead to a decrease in $F^\alpha_{\bar{\nu}\bar{\nu}}(\qv_0,\omega_1)$, where $\bar{\nu}=-\pi T$.
This can be seen for $\alpha=\ch$ in the lower panel of Fig.~\ref{fig:ev_dyn} for $T=0.15$.
In fact, $F^\ch_{\bar{\nu}\bar{\nu}}(\qv_0,\omega_1)$ grows even faster above $U_M(T=0.15)\gtrsim10.5$.

The driving factor must therefore be the impurity vertex function $f$.
The lower panel of Fig.~\ref{fig:ev_dyn} also shows its matrix element $f^\ch_{\bar{\nu}\bar{\nu}\omega_1}$, which indeed shows a steep increase at $U_M(T=0.15)$.
We also verified that the ratio of $F^\ch$ to $f^\ch$ decreases above $U_M(T=0.15)$,
which shows that vertex corrections contribute less and less in the insulating regime.

The Ward identity $\presuper{0\!}\Lambda^\sz=Z^{-1}$ implies that the divergence should also occur in the spin channel.
However, the lower data set in the top panel of Fig.~\ref{fig:ev_dyn} shows that we did not reach sufficiently low temperatures to
achieve $\Re\lambda_\text{max}\lesssim1$.

We note that often the two-particle self-energy $\gamma$ is used to solve the Bethe-Salpeter equation [cf. Appendix, Eq.~\eqref{eq:bsedmft}].
Here we used the impurity vertex function $f$ instead to solve Eq.~\eqref{eq:bsedmftphi}.
This was done because $\gamma$ shows some divergences that do not occur at the Mott transition~\cite{Schaefer16}
and that have also been found in the Hubbard atom~\cite{Thunstroem18}.

\subsection{Fermion-boson response function}\label{sec:fbresponse}
We evaluate the response of the DOS of the half-filled Hubbard model~\eqref{eq:hubbard} to a small shift of the chemical potential $\mu$ or magnetic field $h$.
To this end, we recall that the static limit $\presuper{\infty\!}L$ of the fermion-boson response function is related
via the Ward identities~\eqref{eq:ward:g3statch} and~\eqref{eq:ward:g3statsp} to the response of the Green's function to these fields.
We sum these identities over the momentum $\kv$ and use the DMFT self-consistency condition, $\frac{1}{N}\sum_\kv G_k=g(\nu_n)$, leading to,
\begin{align}
    \presuper{\infty\!}L^{\ch}_{\text{loc}}(\nu_n)=-\frac{dg(\nu_n)}{d\mu},\;\;\;\presuper{\infty\!}L^{\sz}_{\text{loc}}(\nu_n)=-\frac{dg_\uparrow(\nu_n)}{dh},
\end{align}
where $g$ is the impurity Green's function and $\nu_n$ is a fermionic Matsubara frequency.
Since $g$ gives rise to the DOS one can understand $\presuper{\infty\!}L_{\text{loc}}(\nu_n)$ as the response of the DOS to an applied field $\mu$ or $h$, respectively.
At finite temperature we calculate the static limit as $\presuper{\infty\!}L_{\text{loc}}(\nu_n)=L_{\text{loc}}(\nu_n,\omega_0=0)$ according to Sec.~\ref{sec:extrapolation}.

We show in Appendix~\ref{app:ac:w0} that the analytical continuation of $\presuper{\infty\!}L$ can be performed in the same way as for the Matsubara Green's function,
that is, $-\frac{1}{\pi}\Im \presuper{\infty\!}L_{\text{loc}}(\nu+\imath0^+)$ is the retarded fermion-boson response, where $\nu$ is the real energy,
and $-\frac{1}{\pi}\Im g(\nu+\imath0^+)$ is the DOS.
We do not use the maximum entropy method, for reasons explained below,
and instead perform the analytical continuation $\nu_n\rightarrow\nu+\imath0^+$ of $g$ and $\presuper{\infty\!}L_{\text{loc}}$ via Pad\'e approximants~\cite{Vidberg77}.
Similar to Ref.~\cite{Schoett16}, we improved the result for the DOS by averaging the Pad\'e approximants corresponding to a variable number of input points $g(\nu_n)$,
where $32\leq n_\text{max}\leq256$~\footnote{
    We keep the number of Pad\'e coefficients fixed to the number of input points,
    whereas Ref.~\cite{Schoett16} explores more sophisticated options to improve the Pad\'e spectra.}.
In order to obtain an accurate result for $\presuper{\infty\!}L_{\text{loc}}(\nu_n)$ we use the method described in Ref.~\cite{Krien19},
but the numerical error is nevertheless larger than that of $g(\nu_n)$, due to the vertex corrections.
In the following we present Pad\'e results for $\presuper{\infty\!}L_{\text{loc}}(\nu+\imath0^+)$
that were chosen based on qualitative consistency with the DOS and quantitative agreement with sum rules.
\begin{figure}
    \begin{center}
    \includegraphics[width=0.49\textwidth]{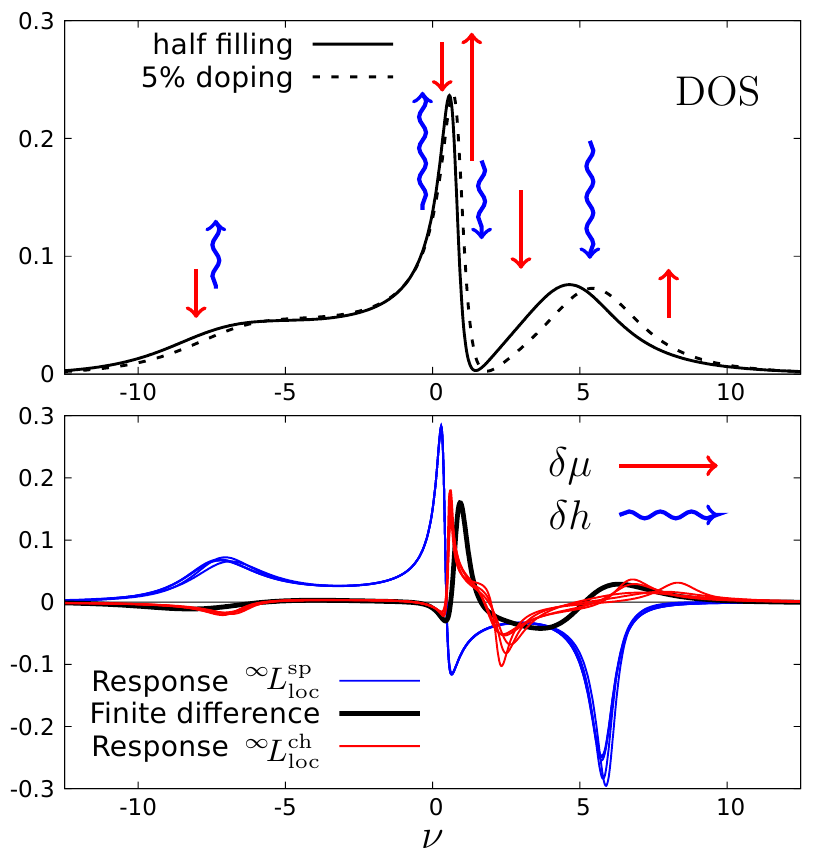}
    \end{center}
    \vspace{-.3cm}
    \caption{\label{fig:fbresponse} (Color online) Response of the DOS to an applied field.
    Top: Black lines show the DOS of the correlated Hubbard model on the triangular lattice in DMFT approximation at half-filling (full) and for $5\%$ doping (dashed).
    Bottom: The thick black line shows the finite difference~\eqref{eq:fdiff}.
    Red lines indicate the charge response, positive (negative) spectral weight implies increase (decrease) of the DOS (see also straight red arrows in the top panel).
    Blue lines and wiggly arrows indicate response to a magnetic field.
    Black-white version, bottom panel: The charge response changes its sign near the maximum of the upper Hubbard band.
    }
\end{figure}

The full black line in the top panel of Fig.~\ref{fig:fbresponse} shows the DOS of the half-filled Hubbard model~\eqref{eq:hubbard}
in the moderately correlated regime $U/t=8$ for a temperature $T/t=0.15$.
The dashed line shows the DOS for the same parameters but at $5\%$ hole-doping.
For the chosen parameters, and due to the particle-hole asymmetry of the triangular lattice, only the right Hubbard band is fully formed.
In this setting some care is required to gain confidence in the qualitative features of the spectrum:
We tested the maximum entropy method to perform the analytical continuation,
however, it persistently predicts a peak just below the Fermi level, also for different default models (not shown, see also Ref.~\cite{Aryanpour06}).
In contrast, our Pad\'e spectra do not predict a two-peak structure near the Fermi level, neither in the result shown in Fig.~\ref{fig:fbresponse},
nor in further tests where we changed the interaction and the temperature.
Furthermore, the Pad\'e spectra in the figure are in good qualitative agreement with numerical renormalization group results (not shown),
which are reliable for small frequency~\cite{Wilson75,Skolimowski19}.

The number of peaks in the DOS is relevant for the number of zero crossings of the charge response $\presuper{\infty\!}L^\ch_{\text{loc}}$.
This is intuitive from the non-interacting limit, where the DOS responds to a change of the chemical potential with a mere shift along the energy axis.
The dashed DOS in the top panel of Fig.~\ref{fig:fbresponse} shows that this is still the main effect in the interacting system.
The doping of $5\%$ corresponds to a shift $|\Delta\mu|/t\sim0.56$ of the chemical potential from its value at half-filling, $\Delta\mu=\mu_{\text{hf}}-\mu_{\text{5\%}}$,
and we can obtain an approximation for the retarded charge response as,
\begin{align}
   -\frac{1}{\pi}\Im \presuper{\infty\!}L^\ch_{\text{loc}}(\nu+\imath0^+)\approx-\frac{\text{DOS}_\text{hf}(\nu)-\text{DOS}_\text{5\%}(\nu)}{\Delta\mu},\label{eq:fdiff}
\end{align}
which is the black line in the bottom panel of Fig.~\ref{fig:fbresponse}.
Positive (negative) spectral weight of this function indicates an increase (decrease) of the DOS upon doping relative to its value at half-filling.
Straight red arrows indicate these trends in the top panel.
The shift of a peak to the right, for example, of the upper Hubbard band,
requires that the charge response has negative weight to the left and positive weight to the right of the peak.
It is therefore plausible that the existence of a peak requires a zero-crossing of the charge response $\presuper{\infty\!}L^\ch$.
On the other hand, the DOS is flat between the quasi-particle peak and the lower Hubbard band and hence the charge response is almost zero in this region.

We discuss our results for the retarded fermion-boson response that we obtained using Pad\'e approximants.
The thin red lines in the bottom panel of Fig.~\ref{fig:fbresponse} show five Pad\'e results that correspond
to different numbers of input points for the Matsubara charge response $\presuper{\infty\!}L^\ch_{\text{loc}}(\nu_n)$.
The spectra shown in the figure were selected on the basis of qualitative agreement with the finite difference~\eqref{eq:fdiff},
where in particular the number and position of the zero crossings are in good agreement.
The bottom panel of Fig.~\ref{fig:fbresponse} also shows the response $-\frac{1}{\pi}\Im\presuper{\infty\!}L^\sz_{\text{loc}}(\nu+\imath0^+)$ to a small magnetic field $h$.
The result is very different from the charge channel, since it discriminates between occupied and unoccupied states.
The DOS is enhanced for $\nu<0$ and suppressed for $\nu>0$, where one should note that the DOS shows the spin-$\uparrow$ states,
whereas for the spin-$\downarrow$ states the sign of $\presuper{\infty\!}L^\sz$ is reversed.
The net effect of the magnetic field is therefore to increase (decrease) the occupancy of spin-$\uparrow (\downarrow)$ states, as expected.
We note that as a benchmark for our Pad\'e spectra we tested two sum rules for the \textit{retarded} fermion-boson response function~\cite{vanLoon18}:
The total integral over $\Im\presuper{\infty\!}L_{\text{loc}}^\alpha(\nu+\imath0^+)$ vanishes~\footnote{
    The integral over the \textit{retarded} response is zero but the integral over the \textit{causal} response yields the total response.
    See also Eq.~\eqref{app:g3contstat} for the relation between retarded and causal response.}, since the spectral weight of the DOS is conserved,
and the integral $\frac{2}{\pi}\int_{-\infty}^{+\infty}d\nu\,n_f(\nu)\Im\presuper{\infty\!}L_{\text{loc}}^\alpha(\nu+\imath0^+)$
yields the total response $\presuper{\infty\!}X^\alpha$, where $n_f$ is the Fermi function.
We verified for $\alpha=\ch,\sz$ that both of these sum rules are indeed satisfied within few percent accuracy.

Lastly, we relate our observations to the Fermi liquid relations that were derived in Sec.~\ref{sec:flparms} for the case of zero temperature.
In this respect one may note that Fig.~\ref{fig:fbresponse} shows that a shift of $\mu$ or $h$ leads to a redistribution of spectral weight across the entire spectrum,
which includes both the response of coherent quasi-particles but also of incoherent states far from the Fermi level.
However, the response of incoherent states plays a quite different role in the charge and spin channels.
For example, the charge response of the upper Hubbard band in Fig.~\ref{fig:fbresponse} roughly cancels upon integration,
while the magnetic response does not change its sign, hence the integral over this region does not cancel.

This explains intuitively the behavior of the total response near the Mott transition at zero temperature:
In this limit the total charge response $\presuper{\infty\!}X^\ch$ is given by the integral
$\frac{2}{\pi}\int_{-\infty}^{0}d\nu\,\Im\presuper{\infty\!}L^\ch_{\text{loc}}(\nu+\imath0^+)$,
where for $Z\rightarrow0$ one integrates merely over the lower Hubbard band,
whose positive and negative contributions to the integral cancel and $\presuper{\infty\!}X^\ch$ vanishes.
In contrast, the corresponding integral over the magnetic response of the lower Hubbard band does not vanish.
For this reason it is plausible that the total magnetic response $\presuper{\infty\!}X^\sz$ remains finite at the Mott transition,
consistent with the results in Fig.~\ref{fig:suscs} for finite temperature.
Furthermore, we showed in Sec.~\ref{sec:z0} that the static charge response $\presuper{\infty\!}L^\ch$
approaches the dynamic response $\presuper{0\!}L^\ch$ [see Eq.~\eqref{eq:fbresponsez0}].
According to the Ward identity~\eqref{eq:ward:g3dyn} $\presuper{0\!}L^\ch_{\text{loc}}$ is given as the frequency derivative of the Green's function, $dg(\nu)/d\nu$,
and therefore in the limit $Z,T\rightarrow0$ the charge response of the Hubbard bands is given by $d\text{DOS}(\nu)/d\nu$.

\section{Summary and Discussion}\label{sec:discussion}
Over the course of this work we have highlighted the important role that two-particle quantities play for the Fermi liquid theory.

An example is the response function $L_{kq}$, which describes the response of individual electronic states with momentum and energy $k=(\kv,\nu)$ to an applied field,
its integral over $k$ yields the susceptibility $X_q$ [see Fig.~\ref{fig:3leg} b)].
Often one is interested in the static homogeneous response, that is, the response to a time-independent homogeneous field.
Therefore, of particular importance is the forward scattering limit $q\rightarrow0$,  
where $q$ comprises the transferred momentum $\qv$ and energy $\omega$ of particle-hole scatterings.
However, in the Fermi liquid the forward scattering limit is ambiguous, since these limits do not commute,
which is a consequence of the pole of weight $Z$ at the Fermi level [cf. Secs.~\ref{sec:flt0:disc} and~\ref{sec:landauparm}].
One refers to the two ambiguous forward scattering limits as the static and the dynamic homogeneous limit, respectively.

One may say that the main line of thought in the derivation~\cite{Landau80,Noziere97,Abrikosov75} of the Fermi liquid theory is to express
the physical static homogeneous limit of several two-particle quantities
in terms of the unphysical dynamic homogeneous limit. The latter is then treated as a free parameter.

For example, closely related to the response function $L$ is the three-leg vertex $\Lambda$ [see Fig.~\ref{fig:3leg}].
Its static limit $\presuper{\infty\!}\Lambda$ can be expressed in terms of the dynamic limit $\presuper{0\!}\Lambda$.
The latter can be calculated at the Fermi level using Ward's identity, $\presuper{0\!}\Lambda=Z^{-1}$,
where $Z$ is assumed to be known, for example, from the experiment.
In turn, the three-leg vertex $\Lambda$ arises from the vertex function $F$ [see Fig.~\ref{fig:3leg} a)].
The static vertex $\presuper{\infty\!}F$ can be expressed in terms of the dynamic vertex $\presuper{0\!}F$,
which defines the Landau parameters $\mathfrak{f}\propto \presuper{0\!}FZ^2$ [see Sec.~\ref{sec:landauparm}], also assumed to be known.
As a result, the quasi-particle weight $Z$, the Landau parameters $\mathfrak{f}$ and the density of states $D(0)$
are the only free parameters of the Fermi liquid theory.

We applied the DMFT approximation to the Bethe-Salpeter equation and arrived at the well-known Fermi liquid relation for the total static response
[see Sec.~\ref{sec:flparms}], $\presuper{\infty\!}X^\alpha=-2\imath D^*(0)/(1+\mathfrak{f}^{\,\alpha})$,
where $D^*(0)$ is the quasi-particle DOS at the Fermi level and $\mathfrak{f}^{\,\ch}, \mathfrak{f}^{\,\sz}$ are the Landau parameters.
In DMFT one routinely calculates the total static response $\presuper{\infty\!}X$.
Thus, when the latter is known, the Landau parameters can be obtained using the exact expression,
\begin{align}
    \mathfrak{f}^{\,\alpha}=-1-2\imath D^*(0)/\presuper{\infty\!}X^\alpha.\label{eq:ffromx_summary}
\end{align}
If $\mathfrak{f}\rightarrow-1$ a Pomeranchuk instability to the phase separation $(\alpha=\ch)$ or to ferromagnetism $(\alpha=\sz)$ occurs.
This criterion is of course equivalent to the divergence of the total response $\presuper{\infty\!}X^\alpha$ in Eq.~\eqref{eq:ffromx_summary}.
We considered the behavior of the Landau parameters at the interaction-driven Mott transition at zero temperature,
where $\mathfrak{f}^{\,\ch}$ diverges $\propto Z^{-2}$, and $\mathfrak{f}^{\,\sz}$ diverges $\propto Z^{-1}$.
Our DMFT calculations at finite temperature confirmed the result for $\mathfrak{f}^{\,\ch}$ [see Sec.~\ref{sec:lparm}],
while our eigenvalue analysis of the Bethe-Salpeter equation showed that the divergences associated to the spin channel are visible only at much lower
temperatures than in the charge channel [cf. Sec.~\ref{sec:eigenvalues}].

Remarkably, in order for the total charge response $\frac{d\langle n\rangle}{d\mu}$ to vanish at the Mott transition,
it is required that the forward scattering amplitude $\presuper{\infty\!}F^\ch_{00}$ diverges $\propto Z^{-1}$ [cf. Eq.~\eqref{eq:xfromf00}].
A further peculiar relation followed for the response function $L$ [see. Sec.~\ref{sec:z0}].
At the Mott transition, and presumably within the entire Mott phase,
the response of the Hubbard bands to the chemical potential is given by the dynamic response, $\presuper{\infty\!}L^{\ch}=\presuper{0\!}L^{\ch}$.
According to the Ward identity it follows that [cf. Sec.~\ref{sec:wardidt0}],
\begin{align}
    \frac{dG_{\kv\nu}}{d\mu}=\frac{dG_{\kv\nu}}{d\nu},\;\;\;\;\;\;(T,Z\rightarrow0),\label{eq:finalresult}
\end{align}    
where $G$ is the causal Green's function, $\mu$ the chemical potential, and $\nu$ the real frequency.
The physical background of Eq.~\eqref{eq:finalresult} is that in the Fermi liquid $\presuper{\infty\!}L^{\ch}$ and $\presuper{0\!}L^{\ch}$
differ by a coherent quasi-particle contribution. The latter vanishes at the Mott transition.

In the spin channel the equivalence of $\presuper{\infty\!}L^{\sz}$ and $\presuper{0\!}L^{\sz}$ does not occur.
In the interacting Fermi liquid both the coherent quasi-particles and incoherent states contribute to the change of the magnetization due to the magnetic field $h$.
At the Mott transition the coherent contribution vanishes, while the incoherent one does not [cf. Sec.~\ref{sec:fbresponse}].
As a consequence, $\presuper{\infty\!}L^{\sz}$ and $\presuper{0\!}L^{\sz}$ are different in the Mott phase.
We verified that $\presuper{\infty\!}L^{\ch}=\presuper{0\!}L^{\ch}$ and $\presuper{\infty\!}L^{\sz}\neq\presuper{0\!}L^{\sz}$
hold in the exactly solvable Hubbard atom at $T=0$ [see Appendix~\ref{app:atom}].

At the two-particle level the Mott transition is characterized by the divergence of the dynamic homogeneous three-leg vertex $\presuper{0\!}\Lambda$~\footnote{
   A relation of the dynamic three-leg vertex to quasi-particle criticality also exists at finite wave-vectors~\cite{Abrahams14,Woelfle17}.}.
This sets this phase transition apart from the more conventional charge and spin Pomeranchuk instabilities,
signaled by divergences of the static vertices $\presuper{\infty\!}\Lambda^\ch=1-\frac{d\Sigma}{d\mu}$
and $\presuper{\infty\!}\Lambda^\sz=1-\frac{d\Sigma}{dh}$.
The latter associate a conjugate field with the respective transition, while $\presuper{0\!}\Lambda=1-\frac{d\Sigma}{d\nu}=Z^{-1}$ does not~\footnote{
    A conjugate field and order parameter for the Mott transition are known in the limit of infinite dimensions~\cite{Zitko15}.}.
It is nevertheless possible to study the Mott transition in a similar way,
by an analysis of the leading eigenvalue of the Bethe-Salpeter equation for a transferred frequency $\omega\neq0$.

In our DMFT calculations this analysis reveals the scattering mechanism that drives the divergence of $\presuper{0\!}\Lambda$ at the Mott transition:
On the Fermi liquid side the divergence is driven by scatterings at many lattice sites,
while in the Mott insulator the scattering amplitude diverges at each site on its own.
This shows how smoothly DMFT captures the breakdown of the Fermi liquid picture at the transition point [Sec.~\ref{sec:eigenvalues}].

It follows that the maximum $d\lambda_\text{max}(U)/dU=0$ of the leading eigenvalue of the Bethe-Salpeter equation~\eqref{eq:bsedmftphi}
may be used to distinguish between the metal and the Mott regime:
In the metal the effect of scatterings at many lattice sites increases with $U$, in the Mott regime this effect decreases [see Fig.~\ref{fig:ev_dyn}].
We find that this criterion is consistent with the drop in the spectral weight at the Fermi level,
which is often used to determine the critical interaction $U_M$ of the transition/crossover [see Fig.~\ref{fig:gtau}].

\section{Conclusions}\label{sec:conclusions}
We have presented a comprehensive analysis of the microscopic Fermi-liquid theory of the single-band Hubbard model and of the Mott-Hubbard transition in the paramagnetic sector.
In particular, we have completely characterized the theory at the two-particle level obtaining complete information about the Landau parameters describing the residual interactions between the heavy quasi-particles with quasi-particle weight $Z$ which vanishes at the Mott transition.

We applied the dynamical mean-field theory (DMFT) approximation to the Bethe-Salpeter equation and derived the Fermi liquid expression,
$\presuper{\infty\!}X=-2\imath D^*(0)/(1+\mathfrak{f})$,
where $\presuper{\infty\!}X$ is the total static homogeneous response function,
$D^*(0)$ the quasi-particle density of states at the Fermi level, and $\mathfrak{f}$ is a Landau parameter.
This well-known result is thus valid in DMFT for an arbitrary lattice dispersion and it allows to calculate the Landau parameters explicitly
from $D^*(0)$ and $\presuper{\infty\!}X$.

Within DMFT the vertex function does not depend on the fermionic momenta. This implies that spatially inhomogeneous deformations of the Fermi surface are not allowed.
As a result, we have only two Landau parameters, $\mathfrak{f}^{\,\ch}$ (symmetric) and $\mathfrak{f}^{\,\sz}$ (anti-symmetric), which correspond to the lowest order ($l=0$) Legendre coefficients in the continuum. 
The two Landau parameters correspond to the two basic Pomeranchuk instabilities of the single-band Hubbard model which can be captured in DMFT,
namely the uniform charge phase separation and ferromagnetic ordering.
In order to obtain Landau parameters of higher order it would be necessary to account for a momentum dependence of the one- and two-particle self-energies.

At the interaction-driven Mott transition at zero temperature we find that the symmetric Landau parameter $\mathfrak{f}^{\,\ch}$ diverges $\propto Z^{-2}$,
where $Z$ is the quasi-particle weight, while the anti-symmetric one $\mathfrak{f}^{\,\sz}$ diverges $\propto Z^{-1}$.
The result for $\mathfrak{f}^{\,\ch}$ is in agreement with the variational Gutzwiller approach to the interaction-driven
metal-insulator transition~\cite{Vollhardt84}. On the other hand, $\mathfrak{f}^{\,\sz}$ remains finite in the Gutzwiller picture,
and the homogeneous spin susceptibility diverges, since this approximation does not capture the effective exchange~\cite{Georges96}, as DMFT does.
We verified the scaling of $\mathfrak{f}^{\,\ch}$ with $Z$ in DMFT calculations for the half-filled Hubbard model on the triangular lattice at finite temperature,
however, we were not able to observe the expected behavior of $\mathfrak{f}^{\,\sz}$.
We suspect that this is due to the correspondence of this Landau parameter to ferromagnetic correlations, which are very weak for intermediate interaction,
and that calculations at zero temperature or at the doping-driven Mott transition for large interaction~\cite{Pruschke03} may confirm our analytical result.

The Ward identity implies the divergence of the dynamic three-leg vertex $\presuper{0\!}\Lambda=Z^{-1}$
and of the dynamic limit of the vertex function $\presuper{0\!}F$ at the critical interaction $U_M$ of the Mott transition.
Our numerical results show that the scattering mechanism that leads to these divergences is non-local on the
Fermi liquid side and local on the Mott side of the transition,
which allows to pinpoint the Mott transition/crossover via an eigenvalue analysis of the Bethe-Salpeter equation,
somewhat reminiscent of the fixpoint analysis of Ref.~\cite{Strand11}.

An exact result of our analysis is that the vanishing of the total charge response $\presuper{\infty\!}X^\ch$ at the Mott transition
requires the static forward scattering vertex $\presuper{\infty\!}F^\ch$ to diverge, as predicted in Ref.~\cite{Chitra01},
and we find that it scales with the quasi-particle weight as $\propto Z^{-1}$. 

It is tempting to connect the divergence of the charge vertex $\presuper{\infty\!}F^\ch$
to the proximity of the Mott insulator to a phase separation instability of the doped Hubbard model,
which can be captured in DMFT by virtue of its frequency-dependent two-particle self-energy~\cite{Yamakawa15,Nourafkan19}.
We speculate that non-local effects beyond DMFT increase the tendency towards phase separation in low dimensional Hubbard models, and in particular in two dimensions.
The calculation of the vertex function across the doping-driven Mott transition thus seems to be an appealing outlook.
However, the finite-doping analysis would require to carefully handle the existence of two solutions leading to the finite-temperature first order Mott transition.

We further discussed the response of individual electronic states to a change of the chemical potential or magnetic field.
The analytical continuation of this response function to the real axis can be done by means of the Ward identity.
We showed that at the Mott transition the charge response of the Hubbard bands is given by the dynamic response,
hence, $\frac{dG}{d\mu}=\frac{dG}{d\nu}$ at the Mott transition, where $G$ is Green's function, $\mu$ the chemical potential, and $\nu$ the real frequency.
We verified that this relation holds in the exactly solvable atomic limit of the Hubbard model.

\acknowledgments
We acknowledge useful comments from D. Vollhardt.
We thank A. Valli for his reading of the manuscript.
F.K. thanks I. Krivenko and D. Guerci for fruitful discussions and
J. Mravlje and R. \v{Z}itko for benchmark results from the NRG,
E.G.C.P. v. L. and M.I.K. acknowledge support from ERC Advanced Grant 338957 FEMTO/NANO. M.C.  acknowledge support from the H2020
Framework Programme, under ERC Advanced GA No. 692670
``FIRSTORM'', MIUR PRIN 2015
(Prot. 2015C5SEJJ001) and SISSA/CNR project "Superconductivity,
Ferroelectricity and Magnetism in bad metals" (Prot. 232/2015).
A.I.L. acknowledges support by the Cluster of Excellence 'Advanced Imaging of Matter' of the Deutsche Forschungsgemeinschaft (DFG) - EXC 2056 - project ID 390715994
and by the North-German Supercomputing Alliance (HLRN) under project number hhp00040.
\appendix
\section{Causal Green's function}\label{app:gf}

We relate the causal Green's function to the retarded, advanced and Matsubara Green's functions.
The causal and the Matsubara Green's function are defined as,
\begin{align}
    G^c_{\kv\sigma}(t)=-\imath\langle T_t c_{\kv\sigma}(t)c^\dagger_{\kv\sigma}(0)\rangle,\\
    G^m_{\kv\sigma}(\tau)=-\langle T_\tau c_{\kv\sigma}(\tau)c^\dagger_{\kv\sigma}(0)\rangle,
\end{align}
respectively, where $t$ is the real time and $\tau$ the imaginary time.
We perform the frequency transforms $G^c(\nu)=\int_{-\infty}^{+\infty} e^{\imath\nu t}G(t)dt$ and $G^m(\nu_n)=\int_{0}^\beta e^{\imath\nu_n\tau}G(\tau)d\tau$,
where $\nu$ and $\nu_n$ are real and Matsubara frequency, respectively.
The spin label $\sigma$ will be dropped. We further define the greater and lesser Green's functions,
\begin{align}
    G^>_\kv(\nu)=&\sum_{ij} w_j |\langle j|c_\kv|i\rangle|^2\delta(\nu-E_i+E_j),\notag\\
    G^<_\kv(\nu)=&\sum_{ij} w_i |\langle j|c_\kv|i\rangle|^2\delta(\nu-E_i+E_j),
\end{align}
where $E_i$ and $|i\rangle$ are the eigenenergies and eigenvectors of the Hubbard model~\eqref{eq:hubbard},
$w_i=\frac{e^{-\beta E_i}}{\mathcal{Z}}$, and $\mathcal{Z}=\sum_i e^{-\beta E_i}$ is the partition sum.

The spectral density can be written as, $S_\kv(\nu)=G^>_\kv(\nu)+G^<_\kv(\nu)$.
We use $S$, $G^>$, and $G^<$ to express the retarded ($r$), advanced ($a$), causal ($c$) and the Matsubara Green's function ($m$),
\begin{align}
    G^{c}_\kv(\nu)=&\int_{-\infty}^\infty\left\{\frac{G^>_\kv(\nu')}{\nu-\nu'+\imath0^+}+\frac{G^<_\kv(\nu')}{\nu-\nu'-\imath0^+}\right\}d\nu'\notag,\\
    G^{m}_\kv(\nu_n)=&\int_{-\infty}^\infty\frac{S_\kv(\nu')}{\imath\nu_n-\nu'}d\nu'\notag,\\
    G^{r/a}_\kv(\nu)=&\int_{-\infty}^\infty\frac{S_\kv(\nu')}{\nu-\nu'\pm\imath0^+}d\nu'.\label{app:retadv}
\end{align}
Here, $0^+$ is a positive infinitesimal real number.
The retarded and advanced Green's functions arise by analytical continuation of $G^m(\imath\nu_n\rightarrow\nu\pm\imath0^+)$ into the upper/lower
complex half-plane, respectively.
The right superscripts $r,a,c,m$ that are used here must not be confused with the left superscript $\mathfrak{r}=|\qv|/\omega$, nor with the channel label $\alpha$. 

We express the causal Green's function in terms of the retarded and advanced ones. Using the identity,
\begin{align}
    \frac{1}{x+\imath0^+}-\frac{1}{x-\imath0^+}=-2\pi\imath\delta(x),\label{app:dirac}
\end{align}
we reformulate $G^c$ in Eq.~\eqref{app:retadv} as,
\begin{align}
    G^{c}_\kv(\nu)=&G^r_\kv(\nu)+2\pi\imath G^<_\kv(\nu)\notag\\
    =& G^r_\kv(\nu)+2\pi\imath S_\kv(\nu)n_f(\nu)\notag\\
    =& \Re G^r_\kv(\nu)+\imath[1-2n_f(\nu)]\Im G^r_\kv(\nu)\notag\\
    =& n_f(-\nu)G^r_\kv(\nu)+n_f(\nu) G^a_\kv(\nu).\label{app:gcgr}
\end{align}
In the first line we used Eq.~\eqref{app:dirac}.
From the first to the second line we used the fluctuation-dissipation theorem, $G^<_\kv(\nu)=e^{-\beta\nu}G^>_\kv(\nu)=S_\kv(\nu)n_f(\nu)$,
where $n_f(\nu)=(e^{\beta\nu}+1)^{-1}$ is the Fermi function. From the second to the third line
we used the relation between the spectral density and the retarded Green's function, $S_\kv(\nu)=-\frac{1}{\pi}\Im G^r_\kv(\nu)$.
In the last step we used $1=n_f(-\nu)+n_f(\nu)$ and $G^a=(G^r)^*$.

Note that the causal Green's function is not positive/negative definite and integrates to
$\frac{1}{2\pi}\int_{-\infty}^{+\infty}d\nu G^c_{\kv\sigma}(\nu)=\imath[\langle n_{\kv\sigma}\rangle-\frac{1}{2}]$,
which can be seen by integrating its Lehmann representation~\eqref{app:retadv}.

\section{Decomposition of the static response}\label{app:decomp}
We derive Eqs.~\eqref{eq:fbvertex} and~\eqref{eq:g3decomp} in the main text.
$k,q$ imply momenta and \textit{real} frequencies, the temperature is zero.

We begin with Eq.~\eqref{eq:bselimits_inv}, which we multiply with the static limit of the bubble $\presuper{\infty\!}G^2_{k'}$ and integrate over $k'$,
then we add $1$ on both sides,
\begin{align}
    &1+\int_{k'}\presuper{\infty\!}F_{kk'}\presuper{\infty\!}G^2_{k'}\\
    =&1+\int_{k'}\presuper{0\!}F_{kk'}\presuper{\infty\!}G^2_{k'}
    +\iint_{k'k''}\presuper{0\!}F_{kk''}R_{k''}\presuper{\infty\!}F_{k''k'}\presuper{\infty\!}G^2_{k'}\notag.
\end{align}
We have dropped the label $\alpha$.
We identify the static three-leg vertex $\presuper{\infty\!}\Lambda_{k}=1+\int_{k'}\presuper{\infty\!}F_{kk'}\presuper{\infty\!}G^2_{k'}$ on the left-hand-side.
In the second term of the right-hand-side we express the static limit $\presuper{\infty\!}G^2$ through the discontinuity $R$ and the dynamic limit $\presuper{0\!}G^2$ [cf. Eq.~\eqref{eq:rdef}],
$\presuper{\infty\!}G^2_{k'}=\presuper{0\!}G^2_{k'}+R_{k'}$, leading to
\begin{align}
    \presuper{\infty\!}\Lambda_{k}=&1+\int_{k'}\presuper{0\!}F_{kk'}\presuper{0\!}G^2_{k'}\notag\\
    +&\int_{k'}\presuper{0\!}F_{kk'}R_{k'}+\iint_{k'k''}\presuper{0\!}F_{kk''}R_{k''}\presuper{\infty\!}F_{k''k'}\presuper{\infty\!}G^2_{k'}\notag.
\end{align}
In the first line we identify the dynamic three-leg vertex $\presuper{0\!}\Lambda_{k}=1+\int_{k'}\presuper{0\!}F_{kk'}\presuper{0\!}G^2_{k'}$,
in the second line we exchange the labels $k'\leftrightarrow k''$ of the double integral and factor out a term $\presuper{0\!}F_{kk'}R_{k'}$,
\begin{align}
    \presuper{\infty\!}\Lambda_{k}=&\presuper{0\!}\Lambda_{k}
    +\int_{k'}\presuper{0\!}F_{kk'}R_{k'}\left(1+\int_{k''}\presuper{\infty\!}F_{k'k''}\presuper{\infty\!}G^2_{k''}\right).\label{app:fbvertex}
\end{align}
The braces yield $\presuper{\infty\!}\Lambda_{k'}$, which leads to Eq.~\eqref{eq:fbvertex} in the main text.%, the Boltzmann equation.

We multiply Eq.~\eqref{app:fbvertex} by $\presuper{\infty\!}G^2_{k}$, this yields the static fermion-boson response function
$\presuper{\infty\!}L_k=\presuper{\infty\!}G^2_{k}\presuper{\infty\!}\Lambda_{k}$ on the left-hand-side,
\begin{align}
    \presuper{\infty\!}L_{k}=&\presuper{\infty\!}G^2_{k}\presuper{0\!}\Lambda_{k}+\presuper{\infty\!}G^2_{k}\int_{k'}\presuper{0\!}F_{kk'}R_{k'}\presuper{\infty\!}\Lambda_{k'}\notag.
\end{align}
We use $\presuper{\infty\!}G^2_{k}=\presuper{0\!}G^2_{k}+R_{k}$ in both terms on the right-hand-side,
\begin{align}
    \presuper{\infty\!}L_{k}=&\presuper{0\!}G^2_{k}\presuper{0\!}\Lambda_{k}+\presuper{0\!}G^2_{k}\int_{k'}\presuper{0\!}F_{kk'}R_{k'}\presuper{\infty\!}\Lambda_{k'}\notag\\
    +&R_{k}\presuper{0\!}\Lambda_{k}+R_{k}\int_{k'}\presuper{0\!}F_{kk'}R_{k'}\presuper{\infty\!}\Lambda_{k'}.\notag
\end{align}
The dynamic fermion-boson response $\presuper{0\!}L_k=\presuper{0\!}G^2_{k}\presuper{0\!}\Lambda_{k}$ arises in the first line,
in the second line we use Eq.~\eqref{eq:fbvertex}, which simply yields $R_{k}\presuper{\infty\!}\Lambda_{k}$, hence,
\begin{align}
    \presuper{\infty\!}L_{k}=&\presuper{0\!}L_{k}+R_{k}\presuper{\infty\!}\Lambda_{k}+\presuper{0\!}G^2_{k}\int_{k'}\presuper{0\!}F_{kk'}R_{k'}\presuper{\infty\!}\Lambda_{k'}\notag
\end{align}
We introduce a trivial integration and factor out a term $R_{k'}\presuper{\infty\!}\Lambda_{k'}$,
\begin{align}
    \presuper{\infty\!}L_{k}=&\presuper{0\!}L_{k}+\int_{k'}\left(\delta_{kk'}
    +\presuper{0\!}G^2_{k}\presuper{0\!}F_{kk'}\right)R_{k'}\presuper{\infty\!}\Lambda_{k'},\label{app:g3decomp}
\end{align}
this is Eq.~\eqref{eq:g3decomp} in the main text. Note that $\delta_{kk'}$ implies a factor $2\pi N$.

\section{The static response in DMFT}\label{app:lindmft}
We derive Eq.~\eqref{eq:g3decompdmft} for the static response $\presuper{\infty\!}L_{k}$ in DMFT.
To this end, we insert the expression for the discontinuity $R_k$ in Eq.~\eqref{eq:rdef} into Eq.~\eqref{app:g3decomp}
(note that the label $\alpha$ is dropped),
\begin{align}
    \presuper{\infty\!}L_{\kv\nu}=&\presuper{0\!}L_{\kv\nu}+\frac{1}{2\pi N}\int_{\nu'}\sum_{\kv'}\left(2\pi N\delta_{\kv\kv'}\delta_{\nu\nu'}
    +\presuper{0\!}G^2_{\kv\nu}\presuper{0\!}F_{\nu\nu'}\right)\notag\\
    \times&\left[-2\pi\imath Z^2\delta(\nu')\delta(\tilde{\varepsilon}_{\kv'}-\mu)\right]\presuper{\infty\!}\Lambda_{\nu'}.\label{app:g3decompexplicit}
\end{align}
Here we have made all energy-momentum dependencies $k=(\kv,\nu)$ and the prefactors of $\int_k=\frac{1}{2\pi N}\int_{\nu}\sum_{\kv}$
and $\delta_{kk'}=2\pi N\delta_{\kv\kv'}\delta_{\nu\nu'}$ explicit. We used that $Z$, $\Lambda$ and $F$ do not depend on $\kv$ (or $\kv'$) in DMFT.
We perform the integration/summation in Eq.~\eqref{app:g3decompexplicit},
\begin{align}
    \presuper{\infty\!}L_{\kv\nu}=&\presuper{0\!}L_{\kv\nu}-2\pi\imath Z^2\delta_{\nu0}\delta(\tilde{\varepsilon}_{\kv}-\mu)\presuper{\infty\!}\Lambda_{0}\label{app:lindmft_intermediate}\\
    -&\imath Z^2D^*(0)\presuper{0\!}G^2_{\kv\nu}\presuper{0\!}F_{\nu0}\presuper{\infty\!}\Lambda_{0},\notag
\end{align}
where we used the definition of the quasi-particle DOS, $D^*(0)=\frac{1}{N}\sum_\kv\delta(\tilde{\varepsilon}_{\kv}-\mu)$.
According to Eqs.~\eqref{eq:lambda0dmft} and~\eqref{eq:xdmft} the static three-leg vertex at the Fermi level can be expressed in terms of the total response,
$\presuper{\infty\!}\Lambda_{0}=\presuper{\infty\!}X[-2\imath Z D^*(0)]^{-1}$. Using this expression in Eq.~\eqref{app:lindmft_intermediate}
and factoring out ${\presuper{\infty\!}X Z}/{2}$ leads to Eq.~\eqref{eq:g3decompdmft}.

\section{Ward identity and analytical continuation of three-point functions}\label{app:ac}
We derive an exact relation between the Matsubara and real axis notation of three-point correlation functions by means of the Ward identity.
For further information see also Refs.~\cite{Oguri01} and~\cite{Eliashberg61}.

Firstly, we note that the last line of Eq.~\eqref{app:gcgr} demonstrates that the causal Green's function $G^c(\nu)$
can be decomposed into two functions that are analytical either in the upper or lower complex half-plane,
$G^c(\nu)=n_f(-\nu)G^r(\nu)+n_f(\nu) G^a(\nu)$.
The analytic regions of $G^r$ and $G^a$ combined cover the entire complex plane $\mathbb{C}$ and their prefactors are given by the Fermi function $n_f$.
$G^r$ and $G^a$ can be obtained from the Matsubara Green's function $G^m(\nu_n)$ by analytical continuation into the upper or lower half-plane.
Eq.~\eqref{app:gcgr} therefore allows to recover $G^c$ from $G^m$.

Our strategy is to find a similar decomposition of the causal fermion-boson response function $L^{c}(\nu,\omega)$ into several component functions,
whose analytic regions cover the entire $\mathbb{C}^2$-space spanned by their two complex arguments.
These component functions are supposed to arise by analytical continuation of the Matsubara correlation function $L^{m}(\nu_n,\omega_m)$.
In principle, this task could be approached from the Lehmann representations of $L^{c}$ and $L^{m}$~\cite{Oguri01,Tagliavini18}, which is however tedious.
We choose a simpler approach here using the Ward identity.

\subsection{Fermion-boson response function}\label{app:ac:g3}
We define the causal fermion-boson response function,
\begin{align}
    &L^{c,\alpha}_{\kv\qv}(t_1,t_2,t_3)=\frac{\imath\langle n\rangle}{2}\sum_\sigma G^c_{\kv\sigma}(t_1-t_2)\delta_\qv\delta_{\alpha,\ch}\label{app:gsusc_real}\\
    +&(\imath)^2\frac{1}{2}\sum_{\sigma\sigma'}s^\alpha_{\sigma'\sigma}
    \left\langle{T_t c^{}_{\kv\sigma}(t_1)c^{\dagger}_{\kv+\qv,\sigma'}(t_2)\rho^\alpha_\qv(t_3)}\right\rangle,\notag
\end{align}
where $s^\alpha$ are the Pauli matrices $(\alpha=\ch,x,y,z)$ and
$\rho^\alpha_\qv=\frac{1}{N}\sum_{\kv}c^{\dagger}_{\kv\sigma}s^\alpha_{\sigma\sigma'}c^{}_{\kv+\qv,\sigma'}$ is the respective density operator,
$\langle n\rangle=\langle n_\uparrow\rangle+\langle n_\downarrow\rangle$.
The correlation function in Eq.~\eqref{app:gsusc_real} depends on three real times $t_i$.
One obtains the Matsubara response $L^{m}$ by replacing $t_i\rightarrow\tau_i$,
$G^c\rightarrow G^m$, and omitting the factor $\imath$ in the first line
and the factor $\imath^2$ in the second line of Eq.~\eqref{app:gsusc_real}.
We note that the term in the first line cancels an uncorrelated part of the charge ($\alpha=\ch$) correlation function.
The (connected) susceptibility is given by $X^{c,\alpha}_\qv(t-t')=\frac{2}{N}\sum_{\kv}L^{c,\alpha}_{\kv\qv}(t',t',t)$.

The transformation of $L^{c}$ in Eq.~\eqref{app:gsusc_real} to the frequency domain is defined as,
\begin{align}
    L^{c}(t_1,t_2,t_3)=\frac{1}{(2\pi)^2}
    \iint\limits_{-\infty}^{\;\;\;+\infty} L^{c}_{\nu\omega}e^{-\imath[\nu t_1-(\nu+\omega)t_2+\omega t_3]}d\nu d\omega.
\end{align}
The analogous transform of $L^{m}$ follows by the replacement $(2\pi)^{-1}\int_{-\infty}^{+\infty}d\nu\rightarrow T\sum_{\nu_n}$, where $\nu_n$ is
a fermionic Matsubara frequency, and likewise for the bosonic frequencies $\omega$ and $\omega_m$.

\subsection{Ward identity}\label{app:ac:ward}
The Ward identity is an exact relation between the response function $L^{c}$ in Eq.~\eqref{app:gsusc_real} and the single-particle Green's function $G^c$.
It arises from the continuity equation $\partial_t\rho^\alpha(t)=\imath[\rho^\alpha(t),H]$ of the density operator $\rho^\alpha$~\cite{Behn78}.
For the Matsubara response $L^{m}$ this derivation is exercised in Ref.~\cite{Krien17},
here we merely state the result for the causal response $L^{c}$ in the homogeneous limit $\qv=\qv_0=\mathbf{0}$,
\begin{align}
    -\omega L^{c,\alpha}_{\kv\nu}(\qv_0,\omega)=G^c_{\kv,\nu+\omega}-&G^c_{\kv\nu}\notag\\
    =n_f(-\nu-\omega)G^r_{\kv,\nu+\omega}+&n_f(\nu+\omega)G^a_{\kv,\nu+\omega}\notag\\
    -n_f(-\nu)G^r_{\kv\nu}-&n_f(\nu)G^a_{\kv\nu}.\label{app:wardidcausal}
\end{align}
Note that the correlation functions in the first line are causal.
From the first to the second line we used Eq.~\eqref{app:gcgr}
to express the causal Green's function $G^c$ through the retarded and advanced Green's functions $G^r$, $G^a$, and the Fermi function $n_f$.
Note that the limit $\omega\rightarrow0$ of Eq.~\eqref{app:wardidcausal} implies Eq.~\eqref{eq:ward:g3dyn} in the main text.

We like to relate the expression in Eq.~\eqref{app:wardidcausal} to the Matsubara response $L^{m}$.
As shown in Appendix A of Ref.~\cite{Krien17}, a similar Ward identity holds for $L^m$,
\begin{align}
    -\imath\omega_m L^{m,\alpha}_{\kv\nu_n}(\qv_0,\omega_m)=G^m_{\kv,\nu_n+\omega_m}-G^m_{\kv\nu_n},\label{app:wardidmatsubara}
\end{align}
which is a relation between Matsubara correlation functions, note however the similarity to Eq.~\eqref{app:wardidcausal}.

\subsection{Analytical continuation}\label{app:ac:ac}
The analytic continuation of the right-hand-side of Eq.~\eqref{app:wardidmatsubara} can be performed into four analytic regions
by replacing $\imath(\nu_n+\omega_m)\rightarrow\nu+\omega\pm\imath0^+$ and $\imath\nu_n\rightarrow\nu\pm\imath0^+$.
On the right-hand-side this gives rise to the retarded and advanced Green's functions $G^r$ and $G^a$, respectively.
We denote the four combinations explicitly as,
\begin{align}
   -\omega L^{rr,\alpha}_{\kv\nu}(\qv_0,\omega)=&G^r_{\kv,\nu+\omega}-G^r_{\kv\nu},\notag\\
   -\omega L^{ra,\alpha}_{\kv\nu}(\qv_0,\omega)=&G^r_{\kv,\nu+\omega}-G^a_{\kv\nu},\notag\\
   -\omega L^{ar,\alpha}_{\kv\nu}(\qv_0,\omega)=&G^a_{\kv,\nu+\omega}-G^r_{\kv\nu},\notag\\
   -\omega L^{aa,\alpha}_{\kv\nu}(\qv_0,\omega)=&G^a_{\kv,\nu+\omega}-G^a_{\kv\nu}.\label{app:fourregions}
\end{align}
We use these expressions to rewrite Eq.~\eqref{app:wardidcausal} as,
\begin{align}
    &-\omega L^{c,\alpha}_{\kv\nu}(\qv_0,\omega)\notag\\
    =&-\omega\left\{n_f(-\nu-\omega)L^{rr,\alpha}_{\kv\nu}(\qv_0,\omega)\right.\notag\\
    &+[n_f(\nu+\omega)+n_f(-\nu)-1]L^{ar,\alpha}_{\kv\nu}(\qv_0,\omega)\notag\\
    &+\left.n_f(\nu)L^{aa,\alpha}_{\kv\nu}(\qv_0,\omega)\right\}.\label{app:g3cont}
\end{align}
We have decomposed the causal response $L^{c}$ into retarded and advanced component functions, $L^{rr}, L^{ar}$, and $L^{aa}$,
which can be readily obtained from the Matsubara response $L^{m}$~\footnote{
    $L^{ra}$ is redundant, since $L^{ra}_{\kv,\nu+\omega}(-\omega)=L^{ar}_{\kv\nu}(\omega)$.}.
We are thus able to recover $L^{c}$ from the latter by analytical continuation.

\subsection{Static homogeneous limit}\label{app:ac:w0}
Strictly speaking, Eq.~\eqref{app:g3cont} can only be used to perform the analytic continuation for $\qv=\qv_0=\mathbf{0}$ and $\omega\neq0$, the dynamic homogeneous limit.
However, it is possible to show that that Eq.~\eqref{app:g3cont} also holds in the static homogeneous limit $\omega=0$.
We demonstrate this here explicitly for the homogeneous magnetic response.

To this end, we assume an infinitesimal magnetic field $\delta h$ along the $z$-axis, the Ward identity in Eq.~\eqref{app:wardidcausal}
can then be written in the transversal channels $\alpha=x,y$ as~\footnote{
The derivation of Eq.~\eqref{app:wardidcausalspinstat} needs to be done from the transversal spin channels:
Bubbles of type $G_{\uparrow} G_{\downarrow}$ are used to construct the transversal magnetic response $L^{x,y}$ from the Bethe-Salpeter equation.
The magnetic field $\delta h$ lifts the degeneracy of the poles of $G_{\uparrow}$ and $G_{\downarrow}$.
Therefore, the limits ${\qv\rightarrow\mathbf{0}}$ and ${\omega\rightarrow0}$ of the bubble $G_{\uparrow}{G_{\downarrow}}$
commute for $\delta h\neq0$, which can be seen easily in the non-interacting case.
Taking the limits ${\qv\rightarrow\mathbf{0}}$, ${\omega\rightarrow0}$ and \textit{subsequently}
the limit $\delta h\rightarrow0$ then leads to the static homogeneous limit,
$\lim\limits_{\qv\rightarrow0}\lim\limits_{\omega\rightarrow0}L^{x,y}_{\kv\nu}(\qv,\omega)$.
Hence, in Eq.~\eqref{app:wardidcausalspin} $\omega$ goes effectively to zero before $\qv$,
which can not be achieved without a symmetry-breaking field [cf. Eq.~\eqref{app:wardidcausal}].
This trick does not work in the channel $\alpha=z$, in this channel the response function $L^z$ is constructed from bubbles of the type
$G_{\sigma}{G_{\sigma}}$, hence $\delta h$ does not lift the degeneracy of the poles.},
\begin{align}
    (2\sigma\delta h-\omega) L^{c,\alpha=x,y}_{\kv\nu}(\qv_0,\omega)=G^c_{\kv,\nu+\omega,-\sigma}-G^c_{\kv\nu\sigma}.\label{app:wardidcausalspin}
\end{align}
We can now safely set $\omega=0$, leading to the static homogeneous limit
$\presuper{\infty\!}L^{c}_{\kv\nu}=\lim\limits_{\qv\rightarrow0}\lim\limits_{\omega\rightarrow0}L^{c}_{\kv\nu}(\qv,\omega)$,
divide by $\delta h$ on both sides, and obtain for $\sigma=\uparrow$,
\begin{align}
    &\presuper{\infty\!}L^{c,\alpha=x,y}_{\kv\nu}=\frac{G^c_{\kv\nu\downarrow}-G^c_{\kv\nu\uparrow}}{2\delta h}\label{app:wardidcausalspinstat}\\
    =&\frac{G^c_{\kv\nu\downarrow}-G^c_{\kv\nu}(h=0)}{2\delta h}-\frac{G^c_{\kv\nu\uparrow}-G^c_{\kv\nu}(h=0)}{2\delta h}\notag\\
    =&-\frac{dG^c_{\kv\nu\uparrow}}{dh}=-n_f(-\nu)\frac{dG^r_{\kv\nu\uparrow}}{dh}-n_f(\nu)\frac{G^a_{\kv\nu\uparrow}}{dh}.\notag
\end{align}
In the second line we added and subtracted Green's function at vanishing field $h=0$, leading to the zero-field derivative
$\frac{dG_\sigma}{dh}=\frac{G_\sigma(\delta h)-G_\sigma(h=0)}{\delta h}$.
In the first step of the last line we used that both spin species respond in opposite ways to the magnetic field,
$\frac{dG_{\uparrow}}{dh}=-\frac{dG_{\downarrow}}{dh}$.
In the last step we used again Eq.~\eqref{app:gcgr}.

An analogous calculation for the Matsubara response $L^{m}$ leads to,
\begin{align}
   \presuper{\infty\!}L^{m,\alpha=x,y}_{\kv\nu_n}=&-\frac{dG^m_{\kv\nu_n\uparrow}}{dh}.\label{app:wardidmatsubaraspinstat}
\end{align}
By rotational invariance Eqs.~\eqref{app:wardidcausalspinstat} and~\eqref{app:wardidmatsubaraspinstat} also hold in the longitudinal spin channel $\alpha=z$.
Furthermore, similar results hold for the charge channel $\alpha=\ch$~\cite{Noziere97},
where one has to replace the magnetic field $h$ by the chemical potential $\mu$.

The analytical continuation of Eq.~\eqref{app:wardidmatsubaraspinstat} is straightforward.
There are only two distinct options, $\imath\nu_n\rightarrow\nu\pm\imath0^+$,
giving rise to the retarded and advanced Green's functions, e.g., $\presuper{\infty\!}L^{rr,\sz}=-\frac{dG^{r}}{dh}$
and $\presuper{\infty\!}L^{aa,\sz}=-\frac{dG^{a}}{dh}$.

Using Eq.~\eqref{app:wardidmatsubaraspinstat} we can write Eq.~\eqref{app:wardidcausalspinstat} as,
\begin{align}
    &\presuper{\infty\!}L^{c,\alpha}_{\kv\nu}=n_f(-\nu)\presuper{\infty\!}L^{rr,\alpha}_{\kv\nu}+n_f(\nu)\presuper{\infty\!}L^{aa,\alpha}_{\kv\nu}.
    \label{app:g3contstat}
\end{align}

We are therefore allowed to divide Eq.~\eqref{app:g3cont} by $-\omega$ and use the result also in the static homogeneous limit $\omega=0$.
We verified from the Lehmann representation of $L^{c}$ and $L^{m}$ of the Hubbard atom [cf. Appendix~\ref{app:atom}]
that Eq.~\eqref{app:g3cont} yields the correct causal response $L^{c}$.
This equation was derived for the homogeneous limit $\qv=\qv_0$ of $L$ but we suspect that it displays the analytical continuation of any fermion-boson response function.

\section{Dynamic limit of the three-leg vertex}\label{app:lambdadyn}
We derive Eq.~\eqref{eq:lambdaward} for the homogeneous three-leg vertex $\Lambda^\alpha_\nu(\qv_0,\omega\neq0)$ in the DMFT approximation from the Ward identity~\eqref{eq:localward}
of the AIM. In this section $\nu$ and $\omega$ are Matsubara frequencies.

Making use of the DMFT self-consistency condition~\eqref{eq:dmftsc}, $g_\nu=\frac{1}{N}\sum_{\kv}G_{\kv\nu}$, one writes Eq.~\eqref{eq:localward} as,
\begin{align}
    &\Sigma_{\nu+\omega}-\Sigma_{\nu}\notag\\
    =&\frac{T}{N}\sum_{\kv'\nu'}\gamma^\alpha_{\nu\nu'\omega}G_{\kv'\nu'}G_{\kv',\nu'+\omega}
    \left[G^{-1}_{\kv'\nu'}-G^{-1}_{\kv',\nu'+\omega}\right]\notag.
\end{align}
In the brackets on the right-hand-side we insert the definition of the DMFT Green's function in Eq.~\eqref{eq:gdmft} and divide both sides by $-\imath\omega$,
\begin{align}
    -&\frac{\Sigma_{\nu+\omega}-\Sigma_{\nu}}{\imath\omega}\label{app:dsigmaladder}\\
    =&\frac{T}{N}\sum_{\kv'\nu'}\gamma^\alpha_{\nu\nu'\omega}G_{\kv'\nu'}G_{\kv',\nu'+\omega}\left[1-\frac{\Sigma_{\nu'+\omega}-\Sigma_{\nu'}}{\imath\omega}\right].\notag
\end{align}
We now consider the Bethe-Salpeter equation for the vertex function,
\begin{align}
    &F^\alpha_{\nu\nu'}(\qv,\omega)\label{eq:bsedmft}\\
    =&\gamma^\alpha_{\nu\nu'\omega}+\frac{T}{N}\sum_{\kv''\nu''}\gamma^\alpha_{\nu\nu''\omega}
    G_{\kv''\nu''}G_{\kv''+\qv,\nu''+\omega}F^\alpha_{\nu''\nu'}(\qv,\omega)\notag,
\end{align}
which is equivalent to Eq.~\eqref{eq:bsedmftphi},~\cite{Hafermann14-2}.
We multiply Eq.~\eqref{eq:bsedmft} with the bubble $G_{k'}G_{k'+q}$, sum over $k'=(\kv',\nu')$,
and evaluate the resulting equation at $q=(\qv_0=\zerov,\omega)$, leading to
\begin{align}
    \frac{T}{N}&\sum_{\kv'\nu'}F^\alpha_{\nu\nu'}(\qv_0,\omega)G_{\kv'\nu'}G_{\kv',\nu'+\omega}\label{app:bsewithlegs}\\
    =&\frac{T}{N}\sum_{\kv'\nu'}\gamma^\alpha_{\nu\nu'\omega}G_{\kv'\nu'}G_{\kv',\nu'+\omega}\notag\\
    \times&\left[1+\frac{T}{N}\sum_{\kv''\nu''}F^\alpha_{\nu'\nu''}(\qv_0,\omega)G_{\kv''\nu''}G_{\kv'',\nu''+\omega}\right].\notag
\end{align}
In the steps leading to Eq.~\eqref{app:bsewithlegs} the summation labels $\nu'$ and $\nu''$ on the right-hand-side were exchanged.
By comparison of Eqs.~\eqref{app:dsigmaladder} and~\eqref{app:bsewithlegs} we find that they actually express the same integral equation.
We can therefore identify,
\begin{align}
    -\frac{\Sigma_{\nu+\omega}-\Sigma_{\nu}}{\imath\omega}=\frac{T}{N}\sum_{\kv'\nu'}F^\alpha_{\nu\nu'}(\qv_0,\omega)G_{\kv'\nu'}G_{\kv',\nu'+\omega}.
\end{align}
Adding $1$ on both sides and using the definition of the three-leg vertex in Eq.~\eqref{eq:lambda_dmft} we arrive at Eq.~\eqref{eq:lambdaward}.

\section{Hubbard atom}\label{app:atom}
We derive the static and dynamic limits of the response function $L$ for the Hubbard atom with Hamiltonian
$H=U n_\up n_\dn -\mu (n_\up+n_\dn) - h (n_\up - n_\dn)$.

\subsection{Correlation functions}
Using the basis set $\{|0\rangle,|\uparrow\rangle,|\downarrow\rangle,|\updownarrow\rangle\}$
we can calculate the causal Green's function $G^c$ using the Lehmann representation in Sec.~\ref{app:gf} (we drop the label $c$),
\begin{align}
    &G_\sigma(\nu)=\frac{1}{\mathcal{Z}}\left[\frac{e^{-\beta\mu}}{\nu+\mu+\sigma h+\imath0^+}+\frac{e^{-\beta\sigma h}}{\nu-U+\mu+\sigma h+\imath0^+}\right.\notag\\
    &+\left.\frac{e^{+\beta\sigma h}}{\nu+\mu+\sigma h-\imath0^+}+\frac{e^{-\beta(U-\mu)}}{\nu-U+\mu+\sigma h-\imath0^+}\right],
\end{align}
where $\mathcal{Z}=e^{-h\beta}+e^{+h\beta}+e^{-\mu\beta}+e^{-(U-\mu)\beta}$ is the partition function, $\beta=\frac{1}{T}$ the inverse temperature.

The response function $L(\nu,\omega)$ can be calculated from the Lehmann representation of Eq.~\eqref{app:gsusc_real},~\cite{Oguri01}.
However, to evaluate this function at $\omega=0$ (static) and in the limit $\omega\rightarrow0$ (dynamic), which are in general \textit{not} equivalent,
it is much more convenient to use the Ward identities~\eqref{eq:ward:g3statch},~\eqref{eq:ward:g3statsp}, and~\eqref{eq:ward:g3dyn}.
These yield the static limits $\presuper{\infty\!}L^{\ch}$, $\presuper{\infty\!}L^{\sz}$ and the dynamic limit $\presuper{0\!}L^{\ch}=\presuper{0\!}L^{\sz}=\presuper{0\!}L^{}$
as derivatives of Green's function with respect to $\mu$, $h$, and $\nu$, respectively.
For $\mu=\frac{U}{2}$ and $h=0$ we obtain for the dynamic limit,
\begin{align}
    &\presuper{0\!}L^{}(\nu)=-\frac{dG_\sigma(\nu)}{d\nu}\\
    =&\frac{1}{\mathcal{Z}}\left[\frac{e^{-\frac{U}{2}\beta}}{(\nu+\frac{U}{2}+\imath0^+)^2}
    +\frac{1}{(\nu-\frac{U}{2}+\imath0^+)^2}\right.\notag\\
       +&\left.\frac{1}{(\nu+\frac{U}{2}-\imath0^+)^2}
       +\frac{e^{-\frac{U}{2}\beta}}{(\nu-\frac{U}{2}-\imath0^+)^2}\right].\notag
\end{align}

According to Eq.~\eqref{eq:g3decomp} the static limit can be expressed through the dynamic one and a remainder,
$\presuper{\infty\!}L^{\alpha}=\presuper{0\!}L^{}+\mathcal{L}^\alpha$.
In the Hubbard atom we can indeed express the static limit in this way.
We obtain the following remainder functions for the charge and spin channel,
\begin{align}
    \mathcal{L}^\ch(\nu)&=\frac{\beta e^{-\frac{U}{2}\beta}}{\mathcal{Z}}\left[
        \frac{1}{\nu+\frac{U}{2}+\imath0^+}-\frac{1}{\nu-\frac{U}{2}-\imath0^+}\right],\notag\\
    \mathcal{L}^\sz(\nu)&=\frac{\beta}{\mathcal{Z}}\left[
        \frac{1}{\nu-\frac{U}{2}+\imath0^+}-\frac{1}{\nu+\frac{U}{2}-\imath0^+}\right].\label{app:atom_fbresponse}
\end{align}
In the charge channel the remainder $\mathcal{L}^\ch$ vanishes as $\beta\rightarrow\infty$,
in this limit therefore $\presuper{\infty\!}L^{\ch}(\nu)=\presuper{0\!}L^{}(\nu)$, as expected.
Hence, also the charge susceptibility vanishes, $\presuper{\infty\!}X^\ch=\frac{2}{2\pi}\int_{-\infty}^{+\infty}d\nu\presuper{\infty\!}L^{\ch}(\nu)=-\imath2\beta e^{-\beta\frac{U}{2}}\mathcal{Z}^{-1}\rightarrow0$.
The remaining charge response of the Hubbard peaks is given by $\presuper{0\!}L^{}(\nu)$.
It does not lead to a response of the density $\langle n\rangle$,
since the integral $\presuper{0\!}X=\frac{2}{2\pi}\int_{-\infty}^{+\infty}d\nu\presuper{0\!}L^{}(\nu)$ is zero.
Note that $\mathcal{L}^\ch$ does not vanish for $T>0$, where the charge susceptibility is finite.

In the spin channel the remainder $\mathcal{L}^\sz$ diverges $\propto\beta$, which gives rise to
the divergence of the spin susceptibility $\presuper{\infty\!}X^\sz=-\imath2\beta\mathcal{Z}^{-1}$, corresponding to the local moment.
Since $\mathcal{L}^\sz$ is not zero, it can not be the case that $\presuper{\infty\!}L^{\sz}(\nu)$ and $\presuper{0\!}L^{}(\nu)$ coincide.

\subsection{Causal fermion-boson response}
\begin{figure}
    \begin{center}
    \includegraphics[width=0.49\textwidth]{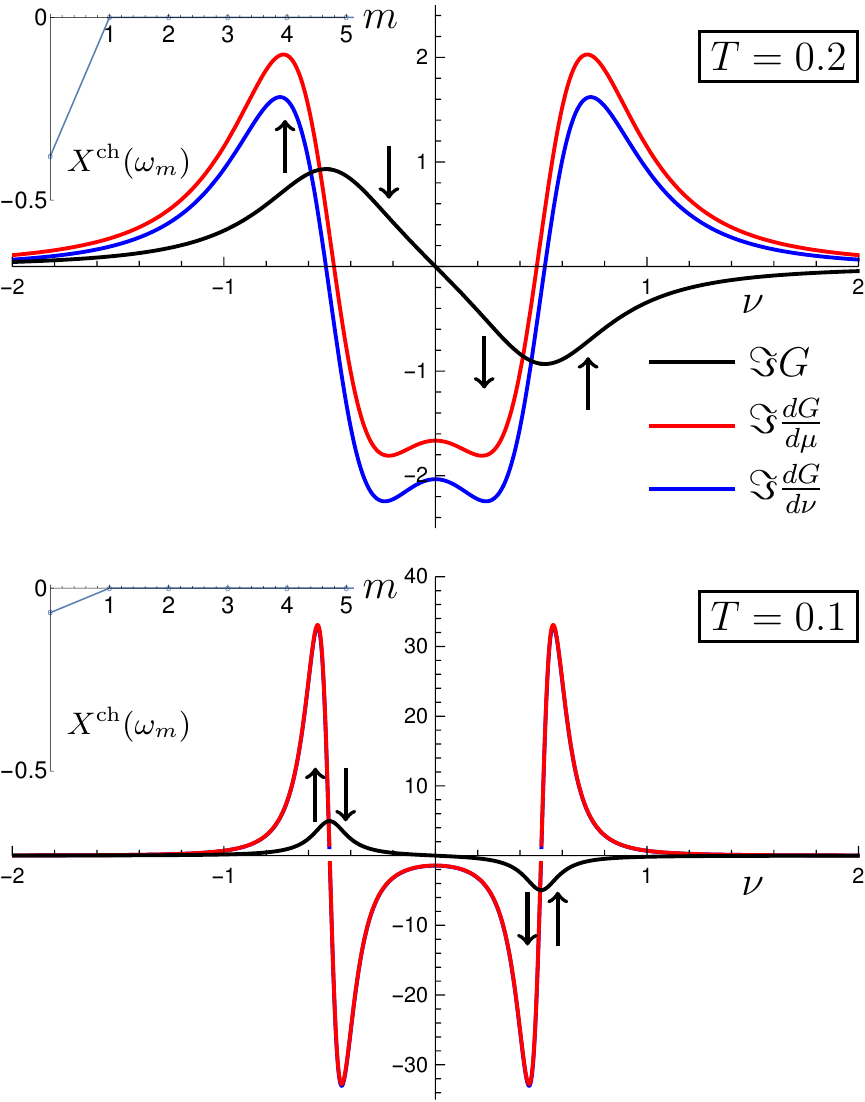}
    \end{center}
    \vspace{-.3cm}
    \caption{\label{fig:alch} (Color online) The causal Green's function $G$ (black), the static charge response $\frac{dG}{d\mu}$ (red),
    and the dynamic response $\frac{dG}{d\nu}$ (blue) of the half-filled Hubbard atom as a function of the real frequency $\nu$.
    Top: $T=0.2$, the static and dynamic response differ appreciably.
    Bottom: $T=0.1$, the limits almost coincide, at the same time the charge susceptibility is suppressed (see insets),
    the latter is given as the integral under the red curve ($\times-\pi^{-1}$).
    Arrows indicate the enhancement/decrease of Green's function according to the red curve due to $\delta\mu>0$.
    }
\end{figure}

\begin{figure}
    \begin{center}
    \includegraphics[width=0.49\textwidth]{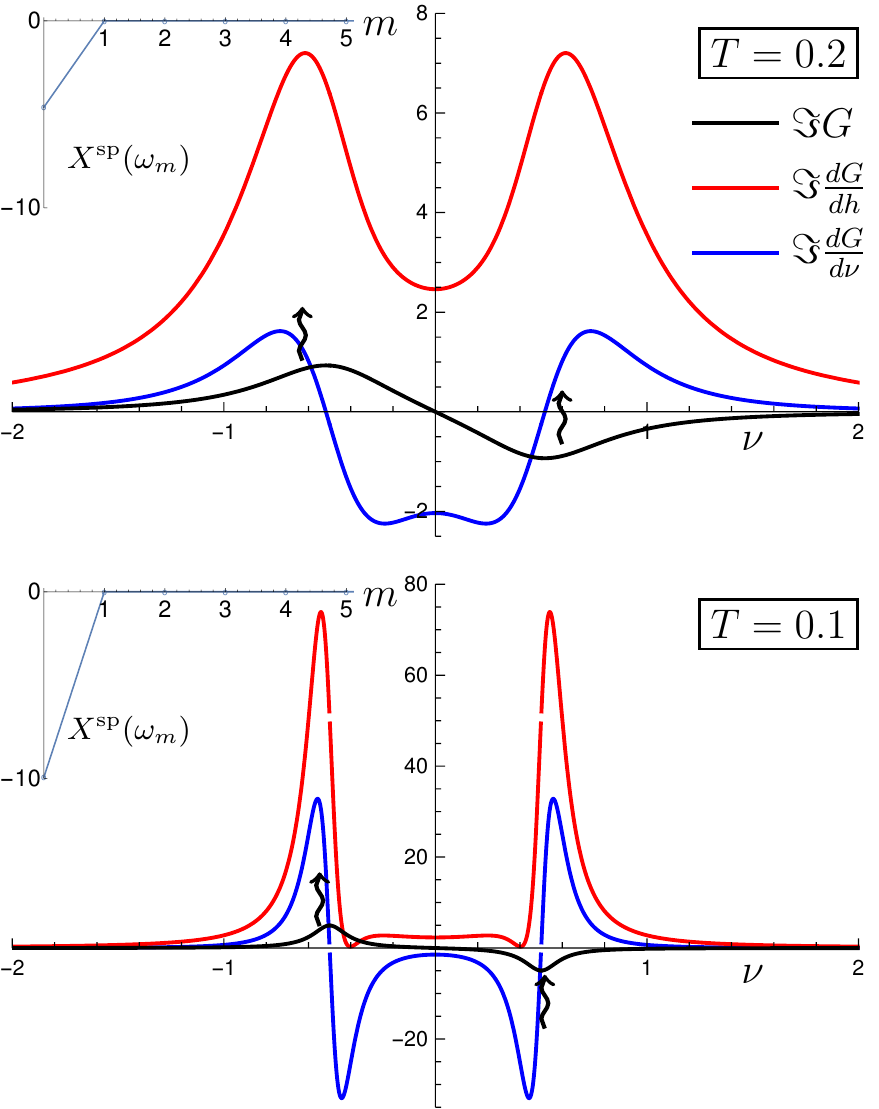}
    \end{center}
    \vspace{-.3cm}
    \caption{\label{fig:alsp} (Color online) The static magnetic response $\frac{dG}{dh}$ (red).
    Note that the causal Green's function (black) and the dynamic response (blue)
    are the same as in Fig.~\ref{fig:alch} for $T=0.2$ (top) and $T=0.1$ (bottom), respectively.
    The difference between the static and dynamic response grows at low temperatures. Wiggly arrows indicate the response to $\delta h>0$.
    The integral under the red curve yields the spin susceptibility ($\times -\pi^{-1}$), see insets, which diverges as $T\rightarrow0$.
    The integral under the blue curve is zero [see text].
    }
\end{figure}

Black lines in Figs.~\ref{fig:alch} and~\ref{fig:alsp} show the causal Green's function $G(\nu)$ for a value of $U=1$
and temperatures $T=0.2$ (top panels) and $T=0.1$ (bottom panels) in units of $U$.
For visibility we use a broadening of $|\eta|=(\pi T)^2$.
We note that the lower Hubbard peak lies below the Fermi level and is therefore hole-like (advanced), giving the peak positive spectral weight,
whereas the upper Hubbard peak is particle-like (retarded) and has negative spectral weight [cf. also Eq.~\eqref{eq:gcgr}].
At half-filling $G_\sigma(\nu)$ integrates to $\frac{1}{2\pi}\int_{-\infty}^{+\infty}G_\sigma(\nu)=\imath[\langle n_\sigma\rangle-\frac{1}{2}]=0$.

The red lines show the imaginary part of the charge response $\frac{dG}{d\mu}=-\presuper{\infty\!}L^{\ch}$ in Fig.~\ref{fig:alch}
and of the magnetic response $\frac{dG}{dh}=-\presuper{\infty\!}L^{\sz}$ in Fig.~\ref{fig:alsp}.
Straight and wiggly arrows indicate where according to $\presuper{\infty\!}L^{\ch}$ and $\presuper{\infty\!}L^{\sz}$ spectral weight of Green's function
is enhanced or suppressed upon a change $\delta\mu$ or $\delta h$ of the respective conjugate field.
Note that a net increase/decrease of spectral weight of the causal Green's function is possible.
In fact, the integral under the red curves yields the static susceptibility,
$\frac{2}{2\pi}\int_{-\infty}^{+\infty}d\nu \presuper{\infty\!}L^{\alpha}(\nu)=\presuper{\infty\!}X^{\alpha}$.
Blue lines indicate the dynamic response function $\frac{dG}{d\nu}=-\presuper{0\!}L^{}$, which is the same for $\alpha=\ch$ and $\alpha=\sz$.
The integral of $\presuper{0\!}L^{}$, the dynamic susceptibility $\presuper{0\!}X$,
is exactly zero because a periodic field, however slowly varying, does not lead to a net change of the particle number or magnetization~\cite{vanLoon18}.

We first discuss the charge response $\presuper{\infty\!}L^{\ch}$ for $T=0.2$ in the top panel of Fig.~\ref{fig:alch}.
$\presuper{\infty\!}L^{\ch}$ changes the spectral weight in such a way that the two Hubbard peaks are effectively
shifted to the left when the chemical potential $\mu$ increases.
At the high temperature $T=0.2$ also the occupation number $\langle n\rangle$ changes due to $\delta\mu$.
Therefore, the integral over $\presuper{\infty\!}L^{\ch}(\nu)$ is finite, representing a net increase of spectral weight due to $\delta\mu>0$.
The resultant charge susceptibility is shown in the inset of the top panel on the Matsubara axis, $\presuper{\infty\!}X^{\ch}$ is marked at $\omega_0=0$.
The top panel of Fig.~\ref{fig:alch} also shows that the static and dynamic charge response $\presuper{\infty\!}L^{\ch}$ (red) and $\presuper{0\!}L^{}$ (blue)
are similar but not equivalent at $T=0.2$. 

We observe the same correlation functions in the lower panel of Fig.~\ref{fig:alch} for a lower temperature $T=0.1$.
Still, $\presuper{\infty\!}L^{\ch}$ indicates a shift of the Hubbard peaks to the left due to $\delta\mu$.
However, charge excitations are suppressed exponentially with decreasing $T$, leading to an almost vanishing charge response $\presuper{\infty\!}X^{\ch}$.

On the same note, $\presuper{\infty\!}L^{\ch}$ and $\presuper{0\!}L^{}$ have become virtually equivalent.
The integral over the former hence (almost) vanishes, since this is exactly the case for the latter.
We note that $\presuper{\infty\!}L^{\ch}=\presuper{0\!}L^{}$ holds exactly at $T=0$.
For its integral we have likewise $\presuper{\infty\!}X^\ch=0$ at $T=0$.

We now turn to the spin channel, whose response function is drawn into Fig.~\ref{fig:alsp}.
A small magnetic field $\delta h$ leads to a shift in spectral weight according to $\presuper{\infty\!}L^{\sz}$,
its effect is qualitatively different from the charge channel.
As indicated by the wiggly arrows, the magnetic field enhances the lower Hubbard peak and suppresses the upper one.
(Note that $G_\uparrow$ is shown, the shift is reversed for $G_\downarrow$.)

We observe that $\presuper{\infty\!}L^{\sz}$ and $\presuper{0\!}L^{}$ are quite different, both at $T=0.2$ and at $T=0.1$.
The analytical result~\eqref{app:atom_fbresponse} shows that they do not become equivalent at $T=0$.
In fact, the spin susceptibility $\presuper{\infty\!}X^{\sz}$ diverges in this limit,
whereas the equivalence of $\presuper{\infty\!}L^{\sz}$ and $\presuper{0\!}L^{}$ would imply a vanishing spin susceptibility.
It follows that $\presuper{\infty\!}L^{\sz}\neq\presuper{0\!}L^{}$.
The integral $\presuper{\infty\!}X^\sz$ represents the response of the Hubbard peaks to the magnetic field
that leads to a net change in the magnetization, $-\delta h\presuper{\infty\!}X^\sz$.

We note that the scenario $\presuper{\infty\!}L^{\ch}=\presuper{0\!}L^{\ch}$ and $\presuper{\infty\!}L^{\sz}\neq\presuper{0\!}L^{\sz}$
at $T=0$ that we find for the Hubbard atom is similar to the one that we found in Sec.~\ref{sec:z0} for the Mott insulator.

\section{Proper definition of the static homogeneous limit}\label{app:limits}
\begin{figure}
    \begin{center}
    \includegraphics[width=0.49\textwidth]{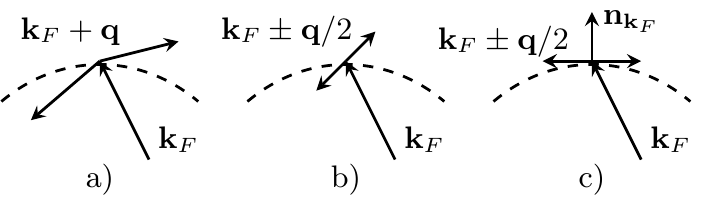}
    \end{center}
    \vspace{-.3cm}
    \caption{\label{fig:pathq} 
    a) In the non-symmetrized notation $G_{k}G_{k+q}$ the vector $\kv_F$ lies on the Fermi surface (dashed line) and $\kv_F+\qv$ inside or outside.
    b) In the symmetrized notation $G_{k+q/2}G_{k-q/2}$ one of the vectors $\kv_F+\qv/2$ and $\kv_F-\qv/2$ lies inside,
    the other outside of the Fermi surface, except for pathological cases.
    c) Pathological case $\qv\perp\mathbf{n}_{\kv_F}$ where even in the limit $\qv\rightarrow\mathbf{0}$
    both $\kv_F+\qv/2$ and $\kv_F-\qv/2$ lie outside of the Fermi surface.}
    \end{figure}

We discuss several technical difficulties that arise in the rigorous evaluation of the static homogeneous limit.
As explained in Sec.~\ref{sec:flt0:disc}, the static homogeneous limit of the the bubble $\presuper{\infty\!}G^2$ describes the propagation of quasi-particle-hole pairs.
In most of this work we use the notation $G_k G_{k+q}$ for the bubble, because it is widespread in the DMFT literature.
However, in this notation the static homogeneous limit seems to be ill-defined,
\begin{align}
    \lim\limits_{\qv\rightarrow\mathbf{0}}\lim\limits_{\omega\rightarrow0}G_{\kv\nu}G_{\kv+\qv,\nu+\omega}
    =\lim\limits_{\qv\rightarrow\mathbf{0}}G_{\kv\nu}G_{\kv+\qv,\nu},\notag
\end{align}
since for some vector $\kv=\kv_F$ on the Fermi surface the vector $\kv_F+\qv$ may lie within or outside of the Fermi surface, depending on the path of $\qv$,
see Fig.~\ref{fig:pathq} a).

We can resolve the ambiguity by adopting the symmetrized notation $G_{k-q/2}G_{k+q/2}$ of Ref.~\cite{Noziere97},
which was used in Sec.~\ref{sec:flt0:disc}, such that $\kv_F-\qv/2$ and $\kv_F+\qv/2$ in general lie inside/outside (outside/inside) of the Fermi surface,
respectively, see Fig.~\ref{fig:pathq} b).
However, even in this notation there arise pathological exceptions, for example, if $\kv_F$ points to some convex region of the Fermi surface
and $\qv\perp\mathbf{n}_{\kv_F}$, where $\mathbf{n}_{\kv_F}$ is the normal of the Fermi surface at $\kv_F$.
In this case $\kv_F-\qv/2$ and $\kv_F+\qv/2$ both lie outside of the Fermi surface, as depicted in Fig.~\ref{fig:pathq} c).
For given $\kv_F$ this problem concerns however only few pathological paths of $\qv$ to zero and it can be resolved by requiring $\qv\not\perp\mathbf{n}_{\kv_F}$.

A further question is raised due to the non-symmetrized notation used in this work for the vertex function $F_{kk'q}$ and the fermion-boson response $L_{kq}$.
For example, in the latter case the relation~\eqref{eq:g3def} to the three-leg vertex $\Lambda_{kq}$ seems problematic, $L_{kq}=G_kG_{k+q}\Lambda_{kq}$,
due to the ambiguous static homogeneous limit of the non-symmetrized bubble $G_kG_{k+q}$.
To resolve the issue, we define the symmetrized static homogeneous limit of $L$ and $F$ as follows,
\begin{align}
    \presuper{\infty\!}L_{k}=&\lim\limits_{\qv\rightarrow\mathbf{0}}\lim\limits_{\omega\rightarrow0}L_{k-q/2,q},\label{app:fblimit_sym}\\
    \presuper{\infty\!}F_{kk'}=&\lim\limits_{\qv\rightarrow\mathbf{0}}\lim\limits_{\omega\rightarrow0}F_{k-q/2,k'-q/2,q},
\end{align}
where we require $\qv\not\perp\mathbf{n}_{\kv_F},\mathbf{n}_{\kv'_F}$ when $\kv=\kv_F$ or $\kv'=\kv'_F$ lie on the Fermi surface.

One should note that $\presuper{\infty\!}L$ can be calculated using the Ward identities~\eqref{eq:ward:g3statch} and~\eqref{eq:ward:g3statsp}.
We derived the latter one in Appendix~\ref{app:ac:w0} by formally taking the non-symmetrized limit $\lim\limits_{\qv\rightarrow\mathbf{0}}\lim\limits_{\omega\rightarrow0}L^\sz_{kq}$,
which leads to the same, unambiguous result as definition~\eqref{app:fblimit_sym}, see also Ref.~\cite{Chubukov18}.
It therefore seems that a symmetrized notation can serve mathematical rigor but does not have physical implications.
 \bibliography{main}

\end{document}